\documentclass{egpubl}
\usepackage{author-version}
\WsPaper
\CGFccby
\usepackage[T1]{fontenc}
\usepackage{dfadobe}  

\usepackage{cite}
\BibtexOrBiblatex

\electronicVersion
\PrintedOrElectronic
\ifpdf \usepackage[pdftex]{graphicx} \pdfcompresslevel=9
\else \usepackage[dvips]{graphicx} \fi

\usepackage{egweblnk}
\usepackage{xspace}
\usepackage{relsize}
\def\code#1{{{\relsize{-1}\texttt{#1}}}\xspace}
\usepackage{amssymb}
\usepackage{relsize}
\usepackage{booktabs}
\usepackage{tabu}
\usepackage{multirow}
\usepackage[capitalise]{cleveref}
\usepackage{tikz}
\usepackage{listings}
\usepackage{stackengine}
\usepackage{algorithm}
\usepackage{algpseudocode}
\usepackage{makecell}
\usepackage{url}
\usepackage{breakurl}

\usepackage{tikz}

\definecolor{ForestGreen}{HTML}{009B55}

\newcommand{\FLIP}{\protect\reflectbox{F}LIP\xspace}

\usepackage{listings}
\newcommand{\INF}{$\infty$}

\newcommand{\tikzpic}[2]{
  \begin{tikzpicture}
  \node[anchor=south west,inner sep=0] (image) at (0,0) {\includegraphics[width=0.32\linewidth]{#1}};
  \node[anchor=south west,inner sep=0] (image) at (0,0) {\includegraphics[width=0.08\linewidth]{#2}};
  \end{tikzpicture}
}

\newcommand{\teasertikz}[2]{
  \begin{tikzpicture}
  \node[anchor=south west,inner sep=0] (image) at (0,0) {\includegraphics[width=0.189\linewidth]{#1}};
  \node[anchor=south west,inner sep=0, text=white] at (0.1,0.1) {#2};
  \end{tikzpicture}
}

\newif\ifsubmission

\title[GPU Volume Rendering with VDB Compression]{GPU Volume Rendering with VDB Compression}
\title[GPU Volume Rendering with Hierarchical Compression Using VDB]{GPU Volume Rendering with Hierarchical Compression Using VDB}
\ifsubmission
\author[Submission ID 1002]{Submission ID: 1002}
\else
\author[S.~Zellmann et al.]
{\parbox{\textwidth}{\vspace{-4em}\centering Stefan~Zellmann$^{1}$\orcid{0000-0003-2880-9090},
    Milan~Jaros$^{2}$\orcid{0000-0003-4630-5339},
    Jefferson~Amstutz$^{3}$\orcid{0000-0001-6002-3739},
    and Ingo~Wald$^{3}$\orcid{0000-0003-0046-713X} 
        }
        \\
{\parbox{\textwidth}{\vspace{-2em}\centering $^1$University of Cologne~$^2$IT4Innovations, VSB–Technical University of Ostrava~$^3$NVIDIA
       }
}
}
\fi

\definecolor{darkgreen}{rgb}{0,0.5,0}
\definecolor{midgreen}{rgb}{0,0.6,0}
\definecolor{lightgreen}{rgb}{0,0.8,0}
\definecolor{darkred}{rgb}{0.6,0,0}

\def\rev#1{#1}

\newif\ifdiff

\ifdiff
\usepackage{soul}

\def\removed#1{\textrm{\color{red}\st{#1}}}
\else

\def\removed#1{}
\fi

\teaser{
  \vspace{-3em}
 \centering
  \teasertikz{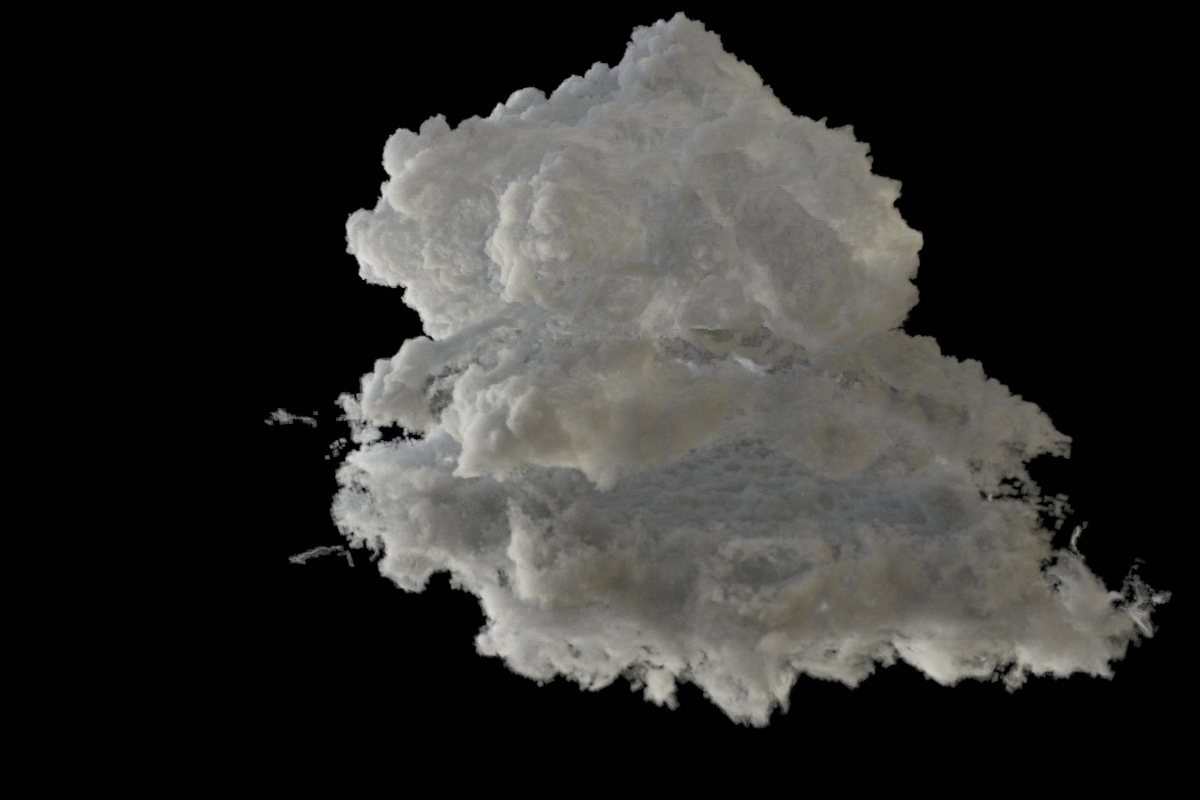}{1:20}
  \teasertikz{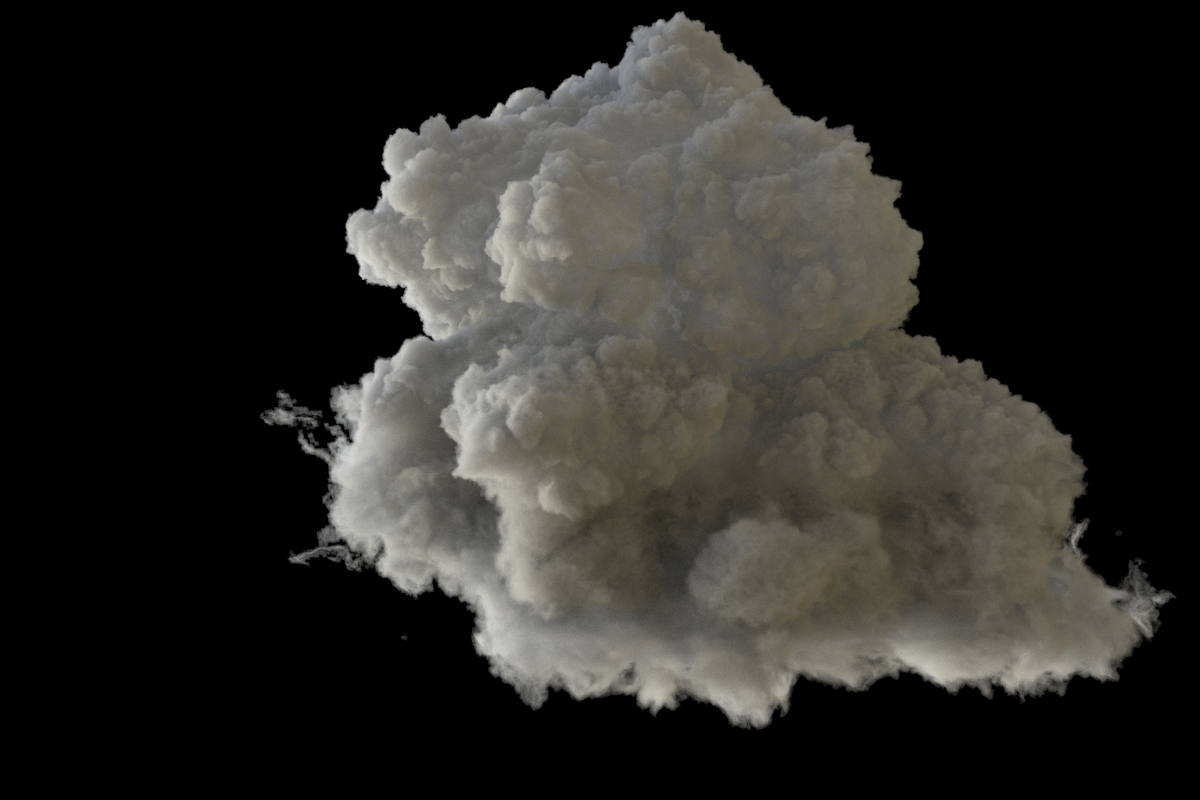}{1:10}
  \teasertikz{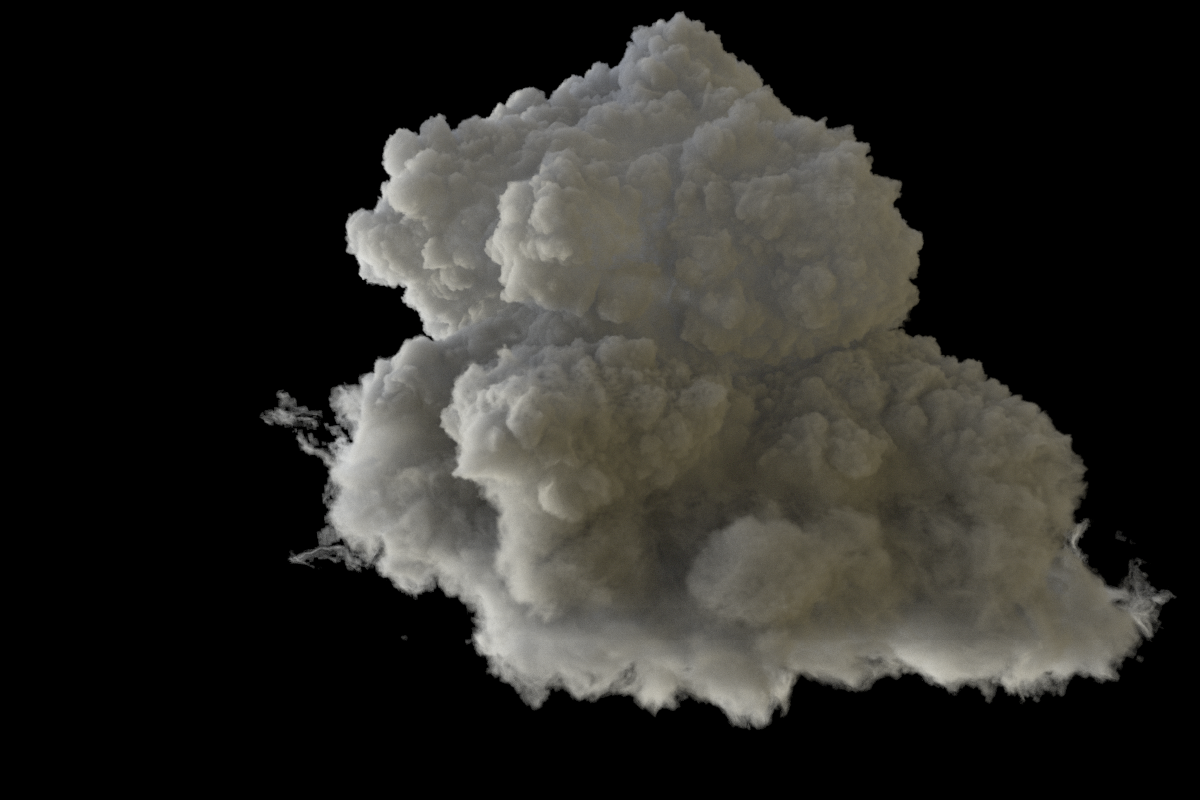}{1:6}
  \teasertikz{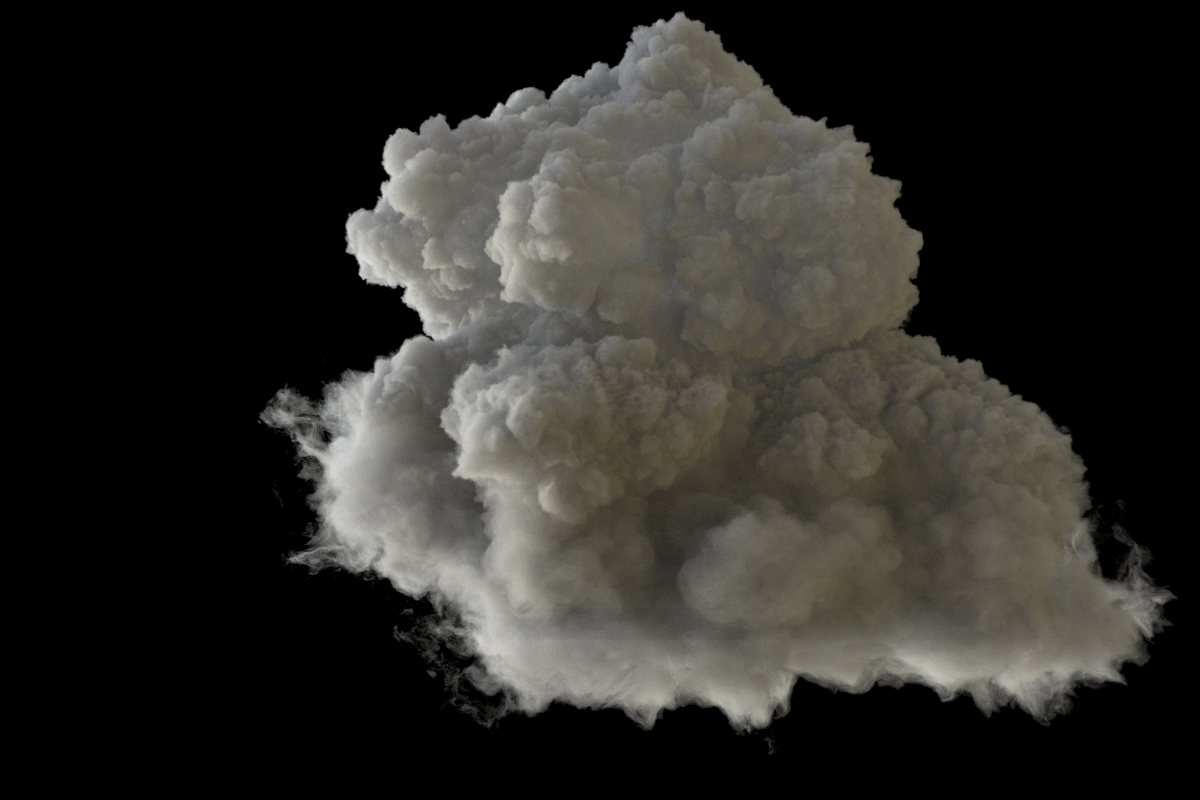}{1:5}
  \teasertikz{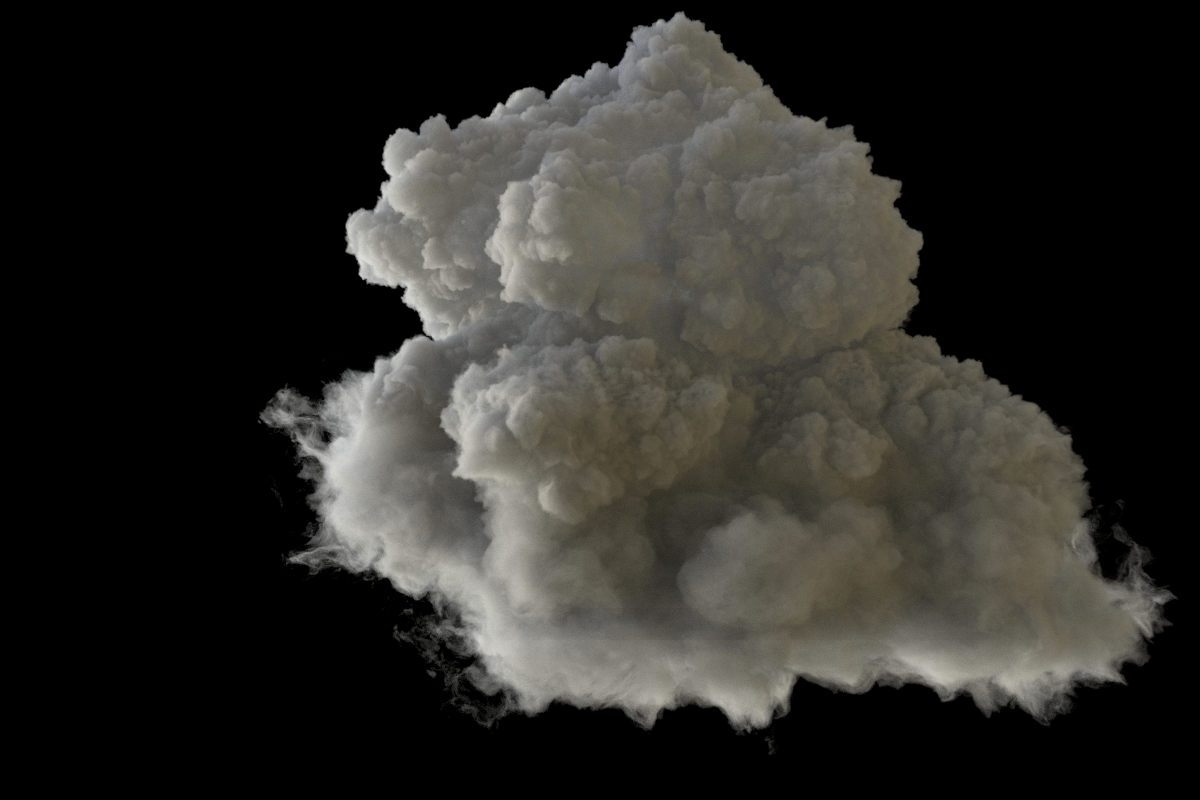}{1:4}
  \vspace{-4pt}

  \teasertikz{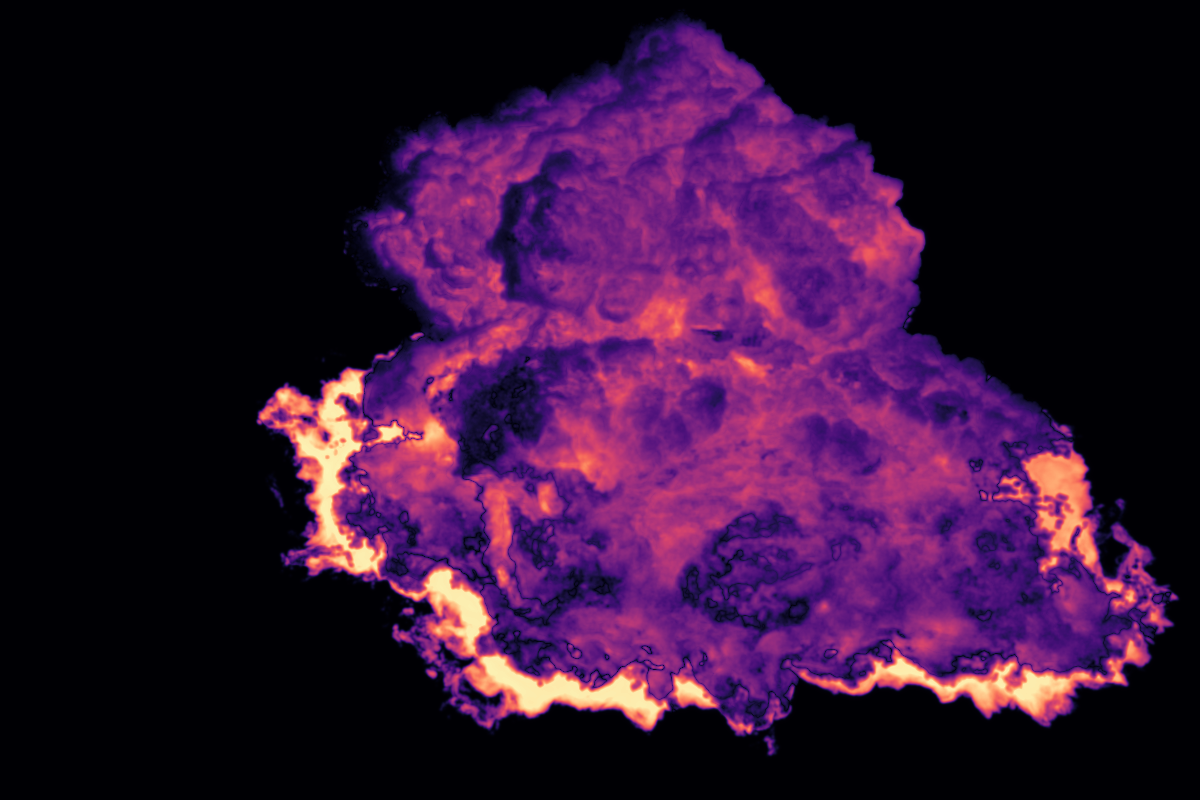}{}
  \teasertikz{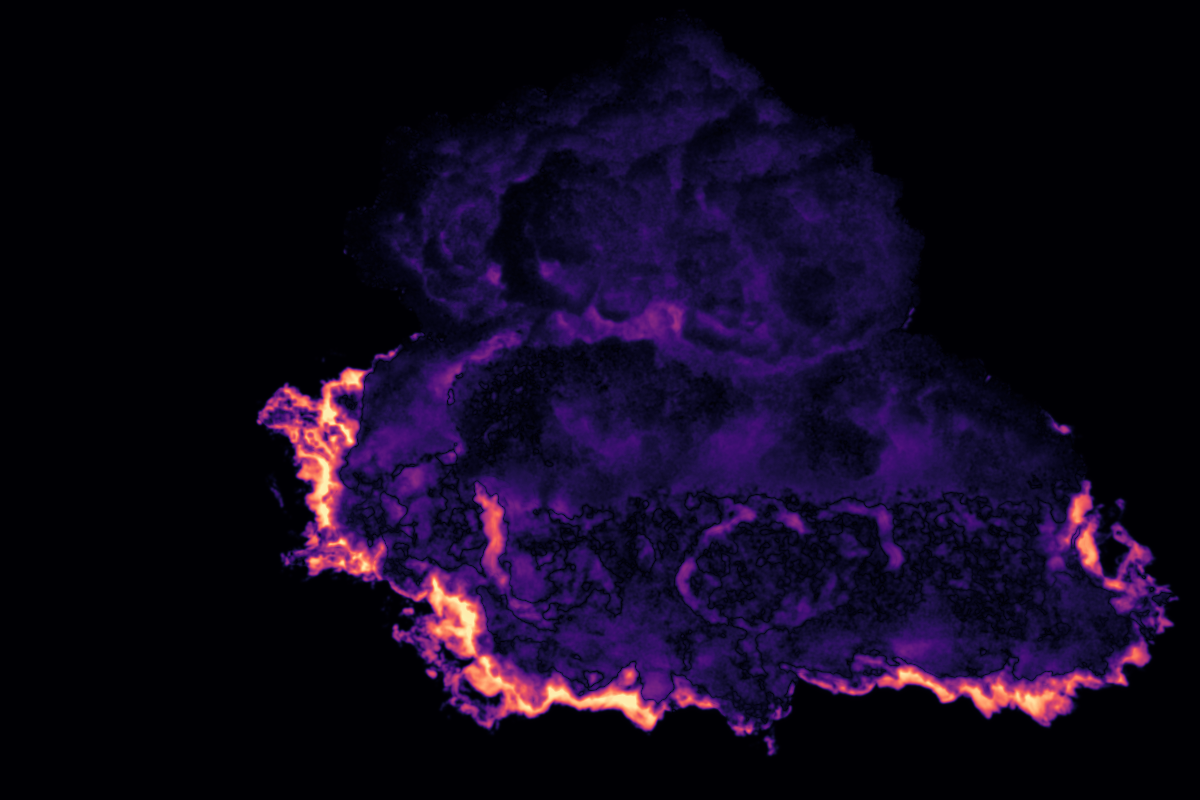}{}
  \teasertikz{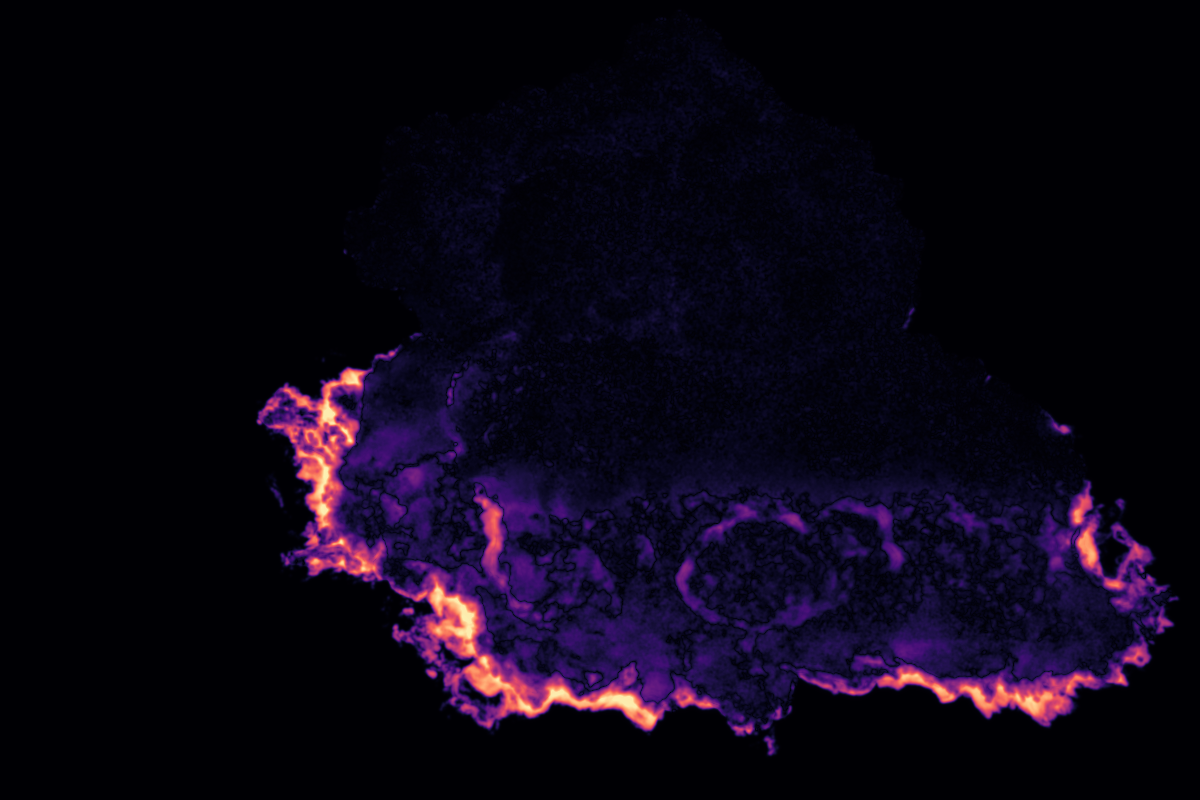}{}
  \teasertikz{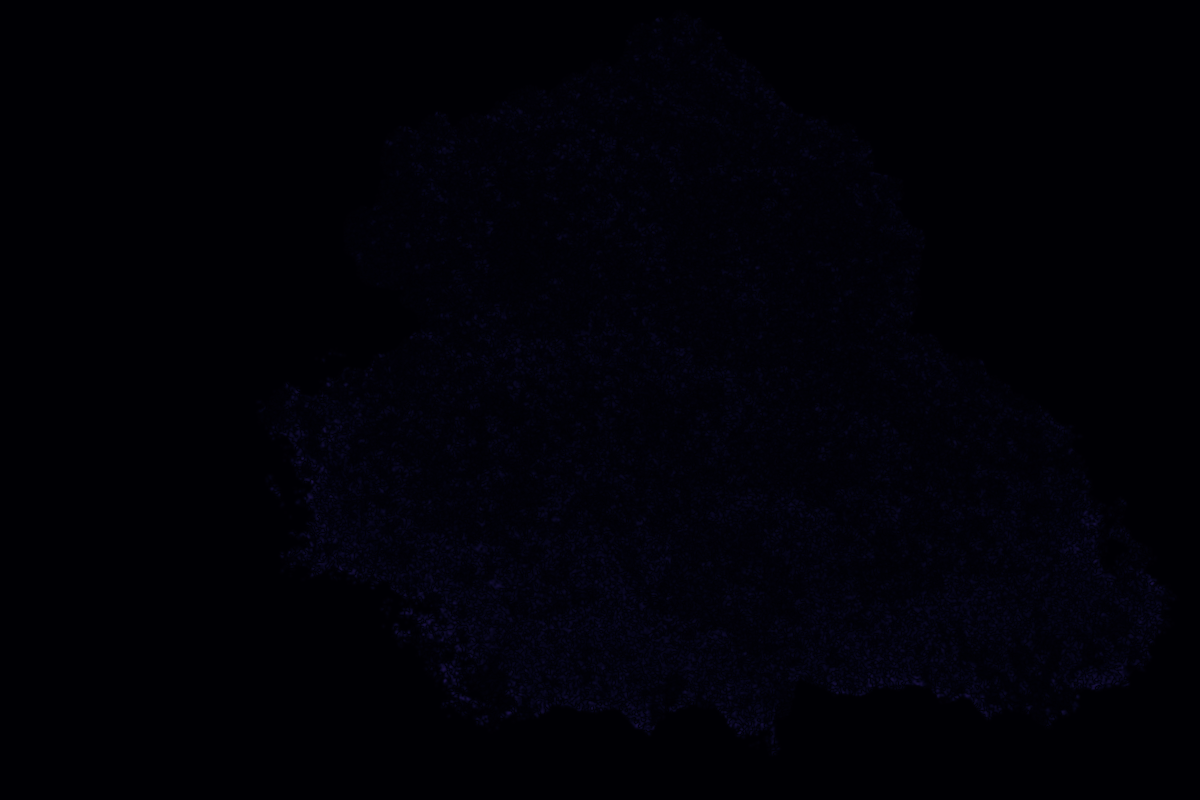}{}
  \teasertikz{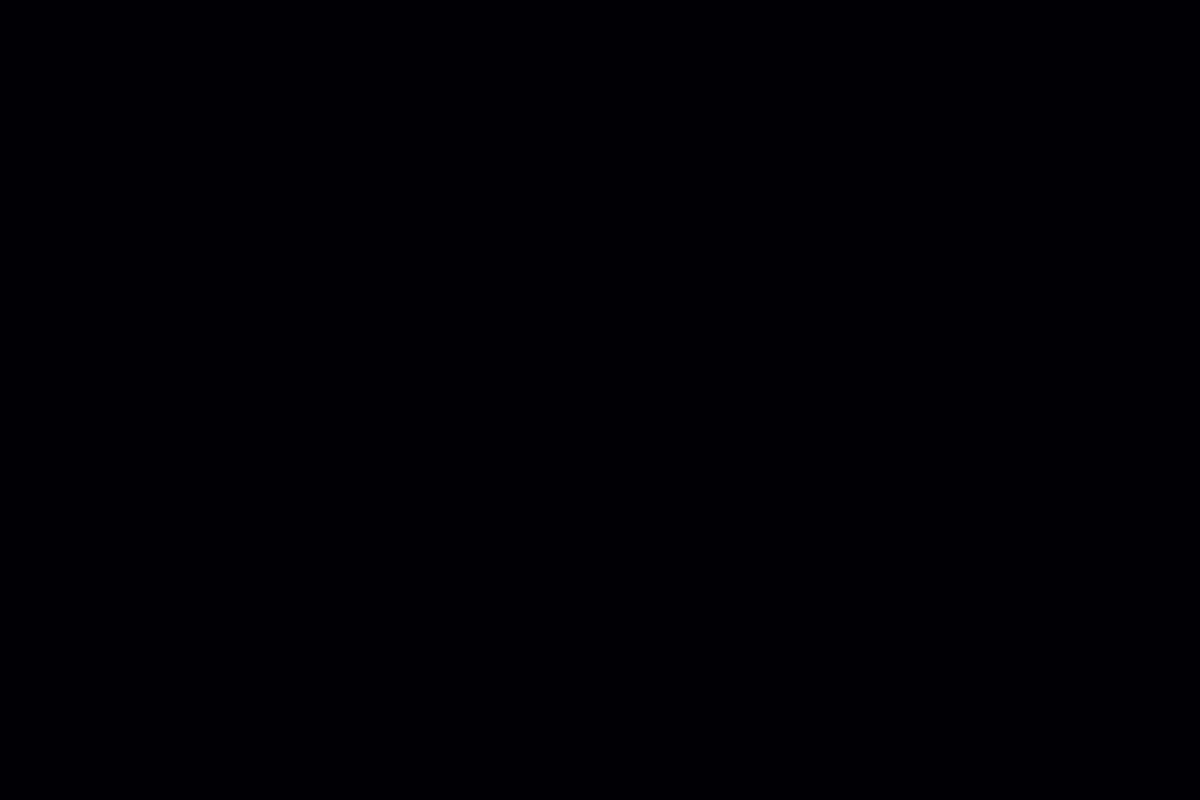}{}
 \vspace{-2em}
 \caption{\label{fig:teaser}%
 WDAS cloud data set compressed at different rates. Top row: path
 traced images of the 2K data set compressed with our proposed method.
 Bottom row: difference images generated with \FLIP comparing the
 compressed ones with renderings of the uncompressed data set.}
\label{fig:teaser}
}

\begin{document}

\maketitle
\begin{abstract}
We propose a compression-based approach to GPU rendering of large volumetric data
using OpenVDB and NanoVDB. We use OpenVDB to create a lossy, fixed-rate
compressed representation of the volume on the host, and use 
NanoVDB to perform fast, low-overhead, and on-the-fly decompression 
during rendering. We show that this approach is fast, works well
even in a (incoherent) Monte Carlo path tracing context, can significantly 
reduce the memory requirements of volume rendering, and
can be used as an almost drop-in replacement into existing 3D texture-based
renderers.
\end{abstract}

\section{Introduction}
\label{sec:intro}
Ray-marched volume rendering in scientific visualization is a well-established
shading style that virtually every 3D visualization application implements.
\rev{One of the research challenges in scientific visualization that remains
is handling the ever-growing size of the data that is visualized.}
GPUs are typically chosen for volume rendering due to their superior memory
bandwidth and texture sampling routines, but this comes at the cost of limiting
(often severely) the size of the data sets that can be rendered. Possible
solutions to that are data parallelism and multi-GPU rendering~\cite{rqs}, or
out-of-core rendering~\cite{sarton}.

Another option to handle the size increase is to use compression. To make use
of standard software for this, one could reach for ZFP~\cite{zfp} with its
fixed-rate compression scheme. As ZFP's fixed-rate algorithm is block-based,
implementing a GPU renderer would require that when blocks are loaded from
memory, many samples are taken across multiple threads to amortize the
decompression costs. For Monte Carlo volume rendering with scattering and
incoherent memory access patterns, this requires using ray wavefronts, barrier
synchronization between bounces, sorting rays for coherency, and other
strategies that result in overly complex control flow. From an engineering
perspective, what is desirable though is random access as known from using 3D
dense textures, so that renderers can devote one GPU thread to each light path.

In this paper we propose to hierarchically encode---and by doing so
compress---the volume data, allowing for per-thread decompression in a GPU
compute kernel. We use the sparse voxel representation (VDB) for that. VDBs,
\rev{so named because they represent volumetric grids that share similarities
with B+trees}~\cite{openvdb},
are popular in production rendering and industry-strength libraries exist that
are optimized for construction and sampling on NVIDIA GPUs. As such, VDBs
are often used for data sets like the one shown in \cref{fig:teaser}.

Our goal is to evaluate if the VDB representation and tools are generally
useful for volume rendering in scientific visualization (sci-vis).
In sci-vis, volumetric representations beyond
structured-regular grids have become more important over the years~\cite{star}.
In practice, this means that volume rendering and sampling libraries need to
maintain multiple data structures for sampling volumes, including
AMR~\cite{wald2021exabrick}, unstructured grids~\cite{bigmesh} or particle
representations~\cite{nate-rbf}. An overarching question we seek to answer
in our research, and which goes beyond the scope of this paper, is if a data
structure like VDB with its efficient and convenient to use GPU implementation
can be used to replace these data types by resampling the data. In this paper,
we explore this question for structured-regular volume representations.

Our main contributions are as follows:
\begin{itemize}
  \item
A fixed-rate compression algorithm with hierarchical encoding implemented with
OpenVDB~\cite{openvdb},
  \item
an interactive volume path tracer decoding such data on-the-fly on the GPU, and
  \item
an evaluation of the framework and algorithm comparing against
dense textures and compression with ZFP.
\end{itemize}
We also integrated our path tracer, which is realized with CUDA and OptiX, with
ANARI. ANARI is an emerging standard for 3D scientific visualization in C++.
This allows us to test the framework using VTK and ParaView.

\section{Related Work}
\label{sec:related}
In this section we review prior work on volume rendering with compression. Our
focus lies on block-based compression, which is often used for direct volume
rendering (DVR). We are specifically interested in the subset of volume
compression algorithms that are used to accelerate renderers, while other
(volume) compression algorithms, e.g., ones that require one to decode the
whole data before sampling, are of minor interest to us. We also look at
related works on hierarchical volume encoding in general, although the majority
of these papers does not focus on compression, but on providing spatial indices
for fast 3D data retrieval or space skipping. We conclude the section with an
overview of VDB grid representations and the OpenVDB and NanoVDB frameworks.

\subsection{Block-Based Volume Compression}
As 3D uniform grids do not adapt to the topology of the underlying data, their
memory footprint increases proportionally when increasing the grid resolution.
Hence, compression has traditionally been the focus of volume rendering-related
works. One way to compress the data is using wavelet transformed blocks in
combination with run-length encoding as proposed by Kim and
Shin~\cite{wavelet}. Back then, volume renderings could not be produced in
real-time, but renderers still required a caching data structure to amortize
decompression costs. Block-based compression remained popular when volume
rendering became interactive due to GPU texture sampling. Schneider and
Westermann~\cite{schneider}, e.g., proposed a block-based algorithm quantizing
the volume across different frequency bands and hierarchically within each
block, and decompressing it in a fragment shader.

In general, block-based compression using the discrete wavelet transform (DWT) or
quantization have been very popular for direct volume rendering. According to
the state-of-the-art report by Rodr\'{i}guez et al.~\cite{star-compressed-vr},
a generic GPU-based compressed direct volume rendering architecture is centered
around encoding blocks during preprocessing, and streaming and decoding those
compressed blocks while rendering (cf.\ Fig.~2 of that report). We also refer the
reader to the state-of-the-art report for a general overview on related work on
compressed volume rendering that goes beyond the scope of our paper.

Block-based data compression for volumetric rendering is still very
popular, which is in parts attributed to the compression framework
ZFP~\cite{zfp}. ZFP has become a de facto standard in the HPC community to
compress 3D volume and other simulation data. We note that ZFP is a
floating-point compression algorithm well applicable beyond 3D volume
rendering, and is in fact able to compress higher dimensional data, while the
scope of our paper is on 3D volume rendering specifically. ZFP's fixed-rate
compression algorithm encodes blocks of size $4 \times 4 \times 4$ containing
floating-point values by first converting them to a common fixed-point format.
This is achieved by factoring out the largest exponent of floating-point values
within each block. The resulting normalized values are converted to a two's
complement fixed-point format. Blocks are then spatially decorrelated by
converting to a different basis based on a tensor product transformation. Doing
so allows the implementation to incorporate different types of transforms,
including DWT, discrete-cosine transform (DCT), and others. The resulting
transform coefficients are then encoded per bit plane. As with the other
block-based compression algorithms, decompression is only efficient when using
caches.

Recent work by Usher et al.~\cite{usher_speculative_2023} ported ZFP to WebGPU.
Their framework decompressed blocks on-the-fly to extract ISO surfaces using
marching cubes. Follow-up work by Dyken et al.~\cite{dyken_prog_iso_ml}
extracted ISO surfaces on-the-fly using ray marching and progressive streaming
of blocks. We are not aware of any other work that specifically uses a scheme
like that applying ZFP to direct volume rendering (DVR); the closest match we
found is the reference renderer that Rapp et al.~\cite{rapp-2022} use for
their evaluation---this renderer compresses a whole set of volume
\emph{samples} for a given perspective frame, though. Implementing a DVR ray
marcher would be conceptually similar to what Dyken et al.\ proposed for ISO
surfaces. We are not aware of any multi-scattering volume renderer using ZFP or
other block-based compression algorithms, but note that caching would be
fundamentally different because of the incoherent memory access patterns used.

\subsection{Hierarchical Volume Representations}
Hierarchical encodings for structured-regular volume data on Cartesian grids
are not necessarily targeted at compression specifically---they usually focus
on aspects such as level-of-detail (LOD) composition, spatial indexing, or to
skip over empty space. In the context of 3D rendering the purpose of that is
often to accelerate the computation, be it by reducing the number of samples
taken, or by representing far away objects with coarser ones. The transition
between hierarchical encoding and compression in general is fluid. We
refer the reader to~\cite{star-compressed-vr} for works on hierarchical
compression for volume rendering. We also note that hierarchical encoding is an
important ingredient to compression in general, e.g., through the use of
Huffman codes, prefix tries, etc., but here specifically focus on hierarchies
that are \emph{spatial} indices.

One popular representation are sparse voxel octrees (SVO)~\cite{laine} that are
often combined with techniques like voxel cone
tracing~\cite{voxel-cone-tracing} for rendering. Such encodings are usually
used to represent surface data as voxels. GigaVoxels~\cite{gigavoxels} is a
framework that implements this paradigm.

Hierarchical encodings for volume data often focus on out-of-core rendering or
empty space skipping.  A comprehensive review on works related to that would be
out of scope for this paper. Instead we refer the reader to the
state-of-the-art report by Sarton et al.~\cite{star}. The encoding closest to
VDBs recently proposed by the literature is BrickTree by Wang et
al.~\cite{bricktree}, which uses a ``wide Octree'' topology built over massive
volume data sets allowing for out-of-core streaming. BrickTree's inner node
encoding uses integer IDs to represent the tree structure, which is similar to
OpenVDB's inner node representation presented below.

\subsection{VDB Grid Representation}
We chose to use the industry standard VDB to implement our method.
VDBs were developed by Museth~\cite{openvdb}, who at that time worked for
DreamWorks Animation. Museth's VDBs are 3D spatial indices for sparse voxel
data with fast, and on average, $O(1)$ random access on modification and
retrieval operations.  The VDB data structure is well-suited for visual effects
rendering of volumetric data from fluid simulation. VDBs are shallow trees with
four levels. In the following we describe the tree layout implemented in
OpenVDB, the open source library resulting from Museth's paper.

In OpenVDB, leaf nodes store a fixed number of voxels (defaulting to $8 \times
8 \times 8$), and inner nodes have a fixed number of child nodes. The root
level has a variable number of children. At each of the levels the VDB data
structure stores a \emph{direct access bit mask} that provides direct random
access to a binary representation of the local topology. These 64-bit values
encode different information depending on the tree level they are on.
Internal nodes, also referred to as tiles, e.g., encode child ID or other
representative values for the whole tile in the bit mask. Leaf nodes use the bit
mask to encode the leaf origin, with $3 \times 20$ bits; \rev{other bits encode
additional information, e.g.,} if the voxels are compressed or quantized. Leaf nodes also
contain a bit mask indicating which voxels are active, and finally a (raw)
pointer to the voxel data itself. This representation allows for fast access
and modification and also out-of-core streaming, yet the data prevails in
address spaces visible to the CPU. OpenVDB provides numerous authoring tools
to modify VDBs at runtime \rev{and is well-established because of its integration
into production renderers like Cycles~\cite{cycles}, and into
visual effects software such as Houdini~\cite{houdini} and Cinema4D~\cite{cinema4d}.}

NanoVDB~\cite{nanovdb} is NVIDIA's linear VDB implementation that does not
depend on CPU address spaces. The whole VDB can be compactly copied using
\code{memcpy} operations and is optimized for retrieving voxel data on the GPU.
Support for modification is rudimentary and essentially limited to altering the
values of voxels that already in the VDB and are active. Museth~\cite{nanovdb}
describes NanoVDB as a ``linear snapshot of an OpenVDB data structure'' that
``explicitly avoids memory pointers''. Tools included with NanoVDB focus on
retrieval more so than on modification, such as 0th to 3rd order interpolation,
gradient computation, as well as converters to and from OpenVDB. The 32-byte
aligned NanoVDB representation is compatible with numerous GPGPU and shading
languages, including CUDA that is used by our implementation.

NeuralVDB by Kim et al.~\cite{neural-vdb} is a recent addition to the VDB
family of frameworks that aim at lossy compression of the voxel data using
multi-layer perceptrons (MLPs). These are trained on the sparse voxel data and
are stored at the lower nodes of the VDB that is otherwise equivalent to
OpenVDB. The paper notes that online random access via inference is too slow
for real-time applications, so the recommended approach of using NeuralVDB is
to decode the neural representation into a regular VDB first.

\section{Dense to Sparse Texture Conversion} 
\label{sec:algo}

In the following, we present an algorithm to hierarchically encode and compress
volumetric data in memory. Our main objectives are efficient decompression with
random access on GPUs, support for single-threaded access from GPU compute
kernels without the requirement to cache any data that is shared with other
threads, and fixed-rate encoding to give us control over the size of the
compressed data. We construct that compressed representation with
OpenVDB~\cite{openvdb}. As OpenVDB is optimized for editing volumes, but not
for random access on the GPU, we then convert the compressed volume with
NanoVDB~\cite{nanovdb}, which gives us an efficient GPU representation that is
linear in memory and optimized for sampling.

\subsection{Fixed-Rate Compression Algorithm} \label{sec:algo}
OpenVDB provides tools and C++ functions to create sparse from dense volumes,
these are however not easily applicable to compression. The integrated tools
require one to identify which value represents \emph{background} (i.e., empty
space). When converting from dense to sparse, only those voxels that are not
background are set and activated and all the others are not. It is easy to
adjust those tools to use thresholding by providing the converter with a
tolerance value. The behavior is hard to control though because the achieved
compression rate is a monotonous, yet not polynomial function of the tolerance
value; increasing the tolerance by just a little might accidentally cull a
majority of voxels of interest.

We propose to use a compression algorithm using histogram and local frequency
analysis of the volume and by that enabling fixed-rate compression. The user
provides a quality parameter in $[0:1]$. As we rely on local frequency
response the algorithm is also well-suited for homogeneous regions with noise
or gradients.

\cref{alg:compression} provides a high-level overview.
The input consists of the volume itself, its spatial extent represented by $W$,
$H$, and $D$ (in voxels), and the user-provided quality value. We define point
sampling routines on the volume itself---without loss of generality, we only
sample the volume at exact (integer) voxel positions though
(\code{volume.value}); as well as routines to activate and set individual
voxels on the VDB (\code{Activate}).
\algdef{SE}[DOWHILE]{Do}{DoWhile}{\algorithmicdo}[1]{\algorithmicwhile\ #1}%
\newcommand{\Break}{\State \textbf{break} }
\begin{algorithm}[ht]
\begin{algorithmic}[1]
\Function{Compress}{$volume$, $W$, $H$, $D$, $quality$}
    \State $hist =$ \Call{computeHistogram}{volume} 
    \State $I =$ \Call{argmax}{$hist$}
    \State $B = volume.valueAt(I)$ \Comment{background value}
    \State
    \State $brickSize =$ \Call{$int3$}{$2**5$}
    \State $numBricks =$ \Call{$int3$}{${W,H,D}/brickSize$}
    \State
    \State \Comment{Compute brick ranges:}{}
    \For{$brickID \in numBricks$}
        \State $lo = brickID*brickSize$
        \State $hi = (brickID+1)*brickSize$
        \State $valueRanges[brickID] = volume.range(lo,hi)$
    \EndFor

    \State $brickRefs =$ \Call{enumBricks()}{}
    \State \Call{SortBy}{$brickRefs$, $valueRanges$, SimilarityFunc($B$)}
    \State
    \State \Comment{Activate important voxels:}{}
    \State $bricksToActivate = numBricks * quality$
    \For{$Brick \in brickRefs$}
        \For{$Voxel \in Brick$}
            \State \Call{Activate}{$Voxel$, $volume.value(Voxel)$}
        \EndFor
        \If{bricksActive++ > bricksToActivate}
            \Break
        \EndIf
    \EndFor
\EndFunction
\end{algorithmic}
\caption{\label{alg:compression}%
Fixed-rate compression and conversion to VDB.
}
\end{algorithm}
The algorithm consists of multiple phases. We assume that a VDB is present and
can be manipulated; the VDB is a 4-level tree (created with OpenVDB) using the
standard $\{3,4,5\}$ tree size configuration. With this configuration,
leaf nodes are of size $2^5$ in each
dimension (i.e., $32 \times 32 \times 32$ voxels); the inner node level above
has $2^4$ children per dimension (i.e., size $16 \times 16 \times 16$), and so
on. The root node level contains as many $2^3$-sized inner nodes as necessary
to cover the volume extent.

In the first phase we compute the histogram for the volume to determine the
background value. As we assume sparsity the value associated with most of the
voxels becomes background; we assume that this value is a good representative
of empty space, i.e., other empty voxels have a value close to this one.

In the second phase we form bricks of size $32 \times 32 \times 32$ to cover
the original volume, so that each brick covers $64$ leaf nodes
of the VDB. We compute value ranges for each brick using the original
volume.

In the third phase we sort those bricks using a list of references, using the
value ranges and a similarity metric. We sort the bricks in descending order
so those whose value contribution is most similar to the background value
come first. We finally iterate over the sorted brick reference list and
activate all the voxels covered by the brick until a predetermined number of
bricks was consumed. By linearly mapping the number of bricks to process to the
user-provided quality value we achieve fixed-rate compression.

To classify voxels as important, we compare their values---and the values
in a local neighborhood---to the background value. We do this based on the brick
decomposition from before, as classifying individual voxels would have an
impractical memory footprint.

We classify bricks as similar to the background value using their value
range; for example, given background value $B$, a brick with value range $[B:B]$
is most similar. Comparing a scalar to a range, we have to
heuristically pick a representative point of that range to compute the
distance of that point to the background value. We propose and evaluate the
following functions:
\begin{eqnarray}
f1 &=& \min(|lo-B|, |hi-B|) \\
f2 &=& \max(|lo-B|, |hi-B|) \\
f3 &=& |(lo+hi)/2-B|.
\end{eqnarray}
$f1$ and $f2$ compute the closest and the farthest distance from the scalar to
the range given by $[lo:hi]$, respectively, while $f3$ computes
the distance to the median of the range. We refer to the metrics as the
closest, farthest and median point-in-range metrics and evaluate their impact
on compression in \cref{sec:eval}.

\subsection{Implementation with OpenVDB and NanoVDB}
To implement the above algorithm we use the C++ library OpenVDB performing the
compression on the CPU. We then convert the result to NanoVDB, which provides a
linear (in memory) representation of the VDB that can be sampled in a shader or
compute kernel but does not allow for arbitrary modification.

\subsubsection{Compression}
We use OpenVDB to implement the fixed-rate compression algorithm. We start from
an empty \code{openvdb::FloatGrid}. In OpenVDB, grids associate trees, the
internal representation of the VDB in memory, with transforms (e.g., voxel to
object space). This internal representation is a \code{Tree4<float, 5, 4, 3>},
i.e., the tree has four levels using the $\{5,4,3\}$ layout as described above.
The tree provides a function \code{addTile(level,Coord(),value)} to set and
activate cells on each tree level.

We start by always activating the extreme (minimum and maximum) corners of the
VDB by just setting the respective voxels to their original value. If we would
not do that because these voxels were identified as background, OpenVDB would
never include them in the tree and the aspect ratio (world bound size) of the
VDB tree would be different from the original volume.

We then create bricks of size $2^5$, compute their scalar ranges, and sort them
by the similarity metric. Given a
user-provided compression rate in percent we can determine the number of bricks
we want to activate. We then iterate over the sorted brick list starting at the one
closest to the background value and activate as many bricks as desired. This is
done by just activating all the voxels (level-0 cells) of that brick. We
finally call \code{tree.prune()} to allow OpenVDB to perform memory
optimizations.

We note that either activating all the cells, or no cells at all, of a brick
can lead to distinct block patterns. Another option would be to decide the
number of voxels we want to activate from the brick based on similarity, too,
e.g.\ that starting at a certain threshold we activate half the number of
voxels but twice the number of bricks. We leave such optimizations as future
work as we would require a (possibly perceptual) metric telling us which voxels
to activate.

Regarding compression performance we note a $1:1$ relationship between
activated voxels and final output size. We also note that activating a whole
tile (\code{addTile()}, see above) on a level higher than leaf node level $0$
results in OpenVDB just setting all the voxels in that tile to the same value,
so there is no additional compression to gain from using hierarchical or
level-of-detail encoding here.

After the VDB was created we convert the OpenVDB grid to NanoVDB using builtin
tools. The NanoVDB representation is linear in memory, can no longer be modify,
but can be copied to the GPU with \code{cudaMemcpy()}.


\subsubsection{On-the-Fly Decompression}
The NanoVDB representation is amenable to sampling on the GPU. NanoVDB provides
C++ utility functions that can be called from host as well as CUDA device code.
This is achieved through accessor and sampler classes. The samplers support
zeroth and first-order interpolation similar to dense textures; samplers using
cubic or higher order interpolation are also available. The samplers use
nearest neighbor or trilinear interpolation on the voxel level (\mbox{level-0} of the
sparse tree) and substitute the background value where no voxels are available.

\section{ANARI Renderer} \label{sec:renderer}
\begin{figure}[tb]
\centering
  \includegraphics[width=0.999\columnwidth]{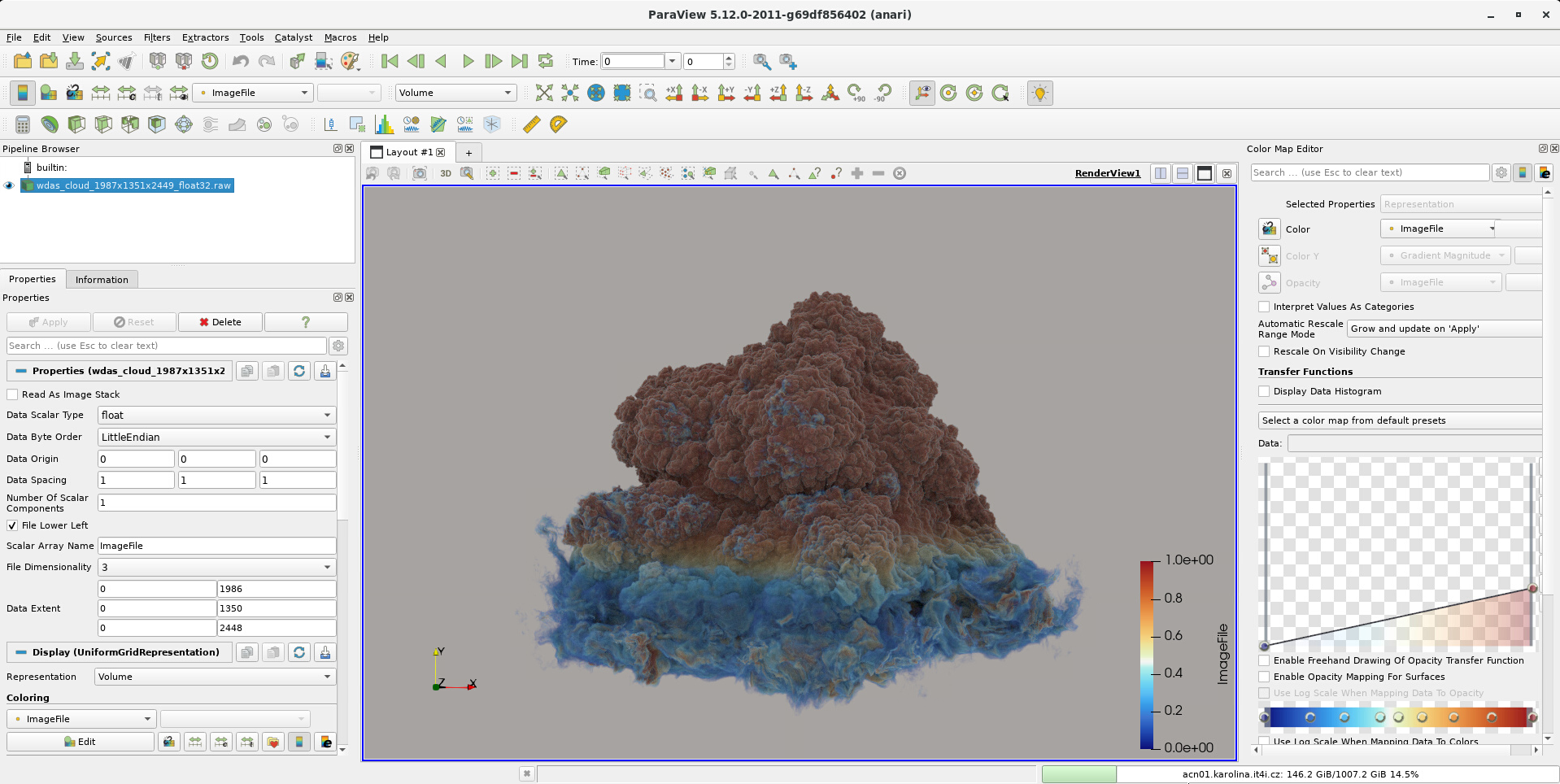}
  \vspace{-2em}
  \caption{\label{fig:cloud-pv}
  WDAS cloud loaded in ParaView using the \code{raw} format encoding. Our
  renderer is integrated using the ANARI interface, structured-regular grids
  are automatically compressed and converted to NanoVDB.
  \vspace{-2em}
  }
\end{figure}
Given our conversion algorithm plus texture sampling routines on the GPU we can
implement a renderer. We chose to give the renderer an ANARI API. ANARI support
has recently been added to ParaView~\cite{dpanari} so our renderer can be
evaluated using standard visualization tools and be compared with other
renderers that also use the ANARI standard (cf. \cref{fig:cloud-pv}).

ANARI standardizes direct volume rendering with structured-regular spatial
fields---the exact type we target with our optimization---but only describes
what data is rendered and not how. We integrate the VDB optimization into
the ANARI renderer Barney~\cite{dpanari}, which is a multi-GPU wavefront path
tracer implementing volume rendering using Monte Carlo free-flight distance
sampling with Woodcock tracking. Multi-GPU rendering is implemented using
Wald et al.'s ray queue cycling method~\cite{rqs}.

We hide our compression algorithm behind an ANSI-C API whose input is the
structured-regular field representation of ANARI and that outputs a NanoVDB
grid. This allows us to hide all the details related to compression behind the
API. The use of VDB is transparent to the user of ANARI, who sets up the data
as though it were an ordinary structured-regular field, with only a ANARI
parameter indicating the data is sparse, and another parameter setting the
desired compression rate. Internally, we replace the device texture object with
a NanoVDB accessor and sampler that we use during sampling from inside Barney's
volume shader.

Barney uses volumetric scattering for shading computations. By default, rays
inside the volume scatter until canceled by Russian Roulette, and radiance is
contributed via image-based lighting through an HDRI light source, or by an
ambient light source constantly illuminating the scene from all directions.
ANARI and Barney provide a render graph so that more than just a single volume
can be present in the scene at a time, and mixed surface and volumetric scenes
are supported. For rendering volumes, Barney implements two operations:
traversal using majorant densities for accelerated free-flight distance
tracking, and random access sampling into the spatial field itself.

Traversal is realized using macrocell grids 
that store min/max ranges of the density the cell represents. These min/max
ranges are used as lookups into an RGBA transfer function; the $\alpha$
component of the transfer function serves as extinction coefficient and is also
used to provide majorants. These majorants are recomputed whenever the transfer
function changes, based on the min/max ranges.

By traversing the majorant grid, the path tracer can adapt the sampling rate to
how homogeneous the density is inside the macrocells. Cells that are empty can
be skipped over; in cells that are homogeneous, the majorant density is closer
to the reconstructed density, which allows the Woodcock tracking algorithm to
take bigger steps so that fewer samples are taken. The grid is traversed using
the 3D digital differential analyzer (3D-DDA) algorithm as
in~\cite{szirmaykalos2011freeps}. When a non-empty cell is encountered,
Woodcock tracking is performed inside that cell. If a valid free-flight
distance was found, traversal stops; otherwise, it continues to the next cell
or until the ray leaves the grid.

With VDB, it is possible to implement a hierarchical version of DDA. Some first
experiments with that showed diminishing returns for that approach. Diminishing
returns of using hierarchical space skipping data structures for volumetric
path tracing in the presence of RGBA transfer functions were also reported
by Zellmann et al.~\cite{zellmann-beyond-exabricks}. Using a uniform grid for traversal also
allows us to compare the sampling performance with structured-regular volumes
using dense textures. Therefore, we opted to use a uniform grid with
non-hierarchical traversal also for the VDB texture representation.

\section{Evaluation} \label{sec:eval}
\begin{table*}[th]
\begin{center}
\setlength{\tabcolsep}{2pt}
\begin{tabular}{ccc}
  \tikzpic{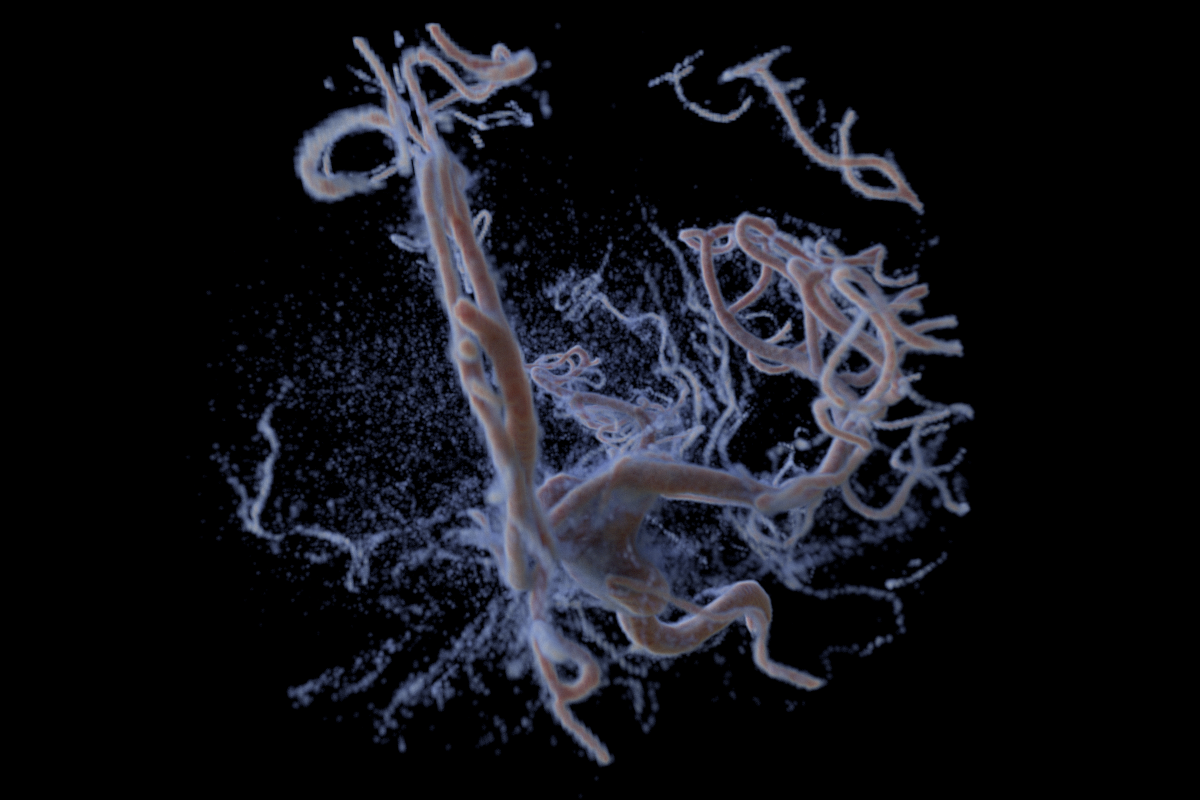}{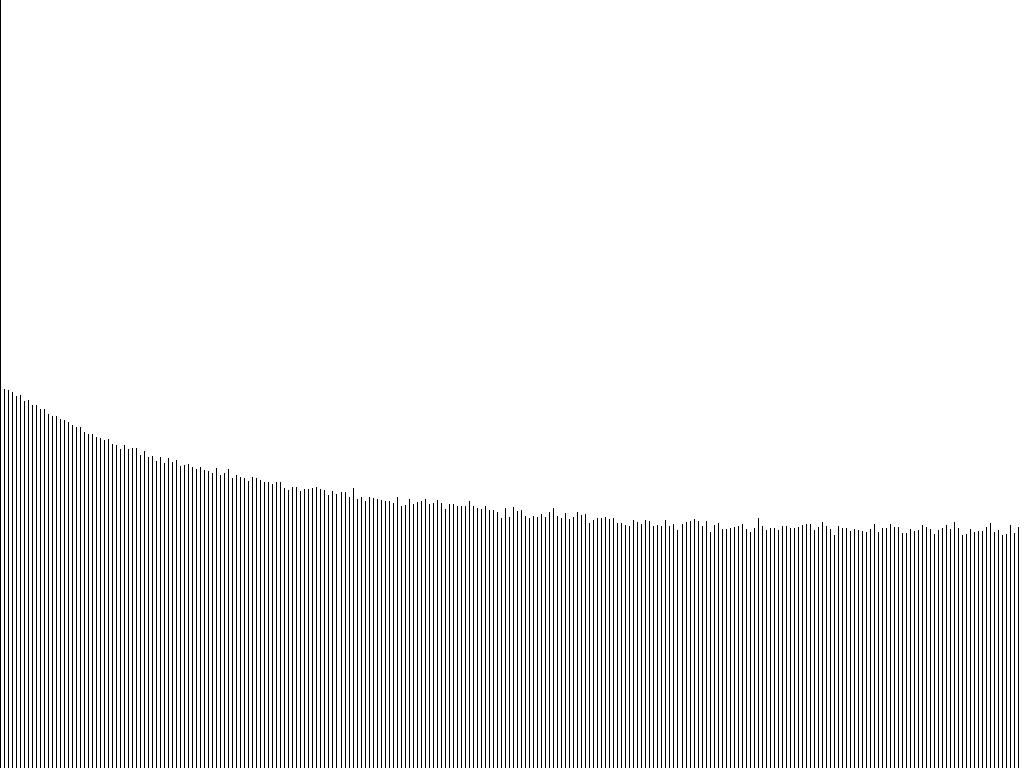}&
  \tikzpic{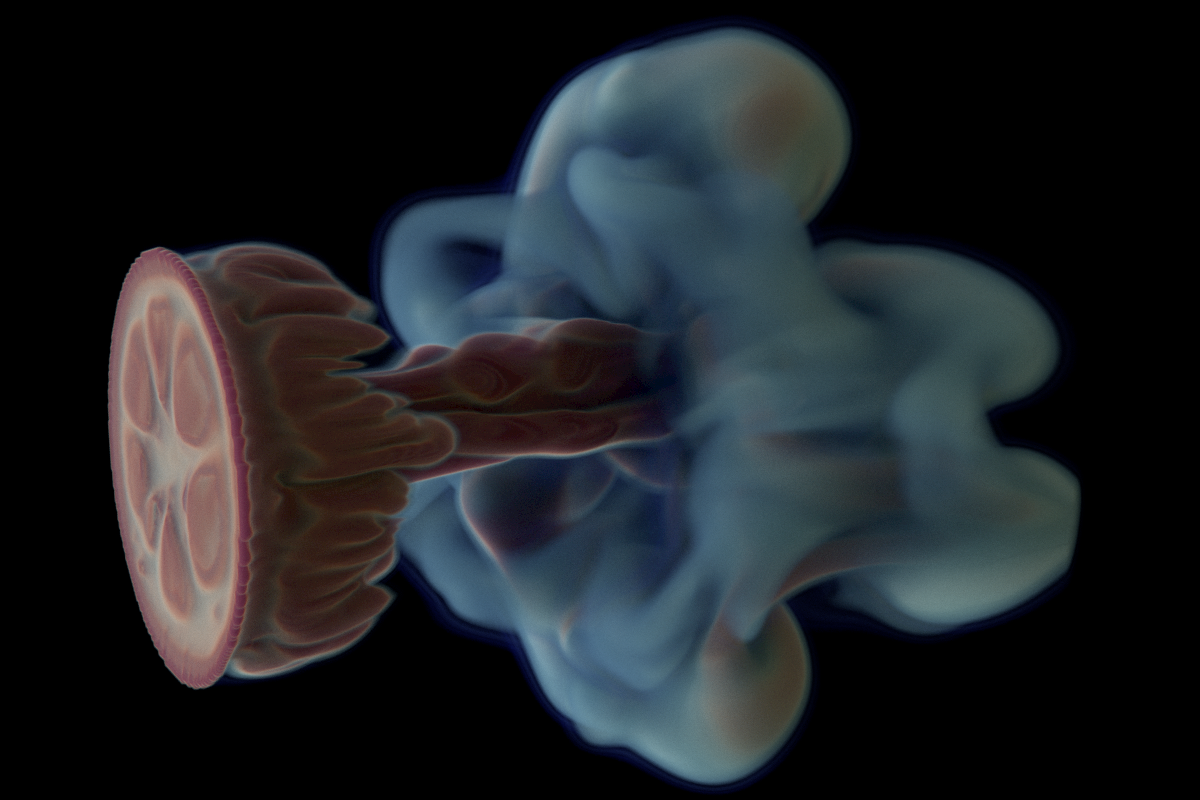}{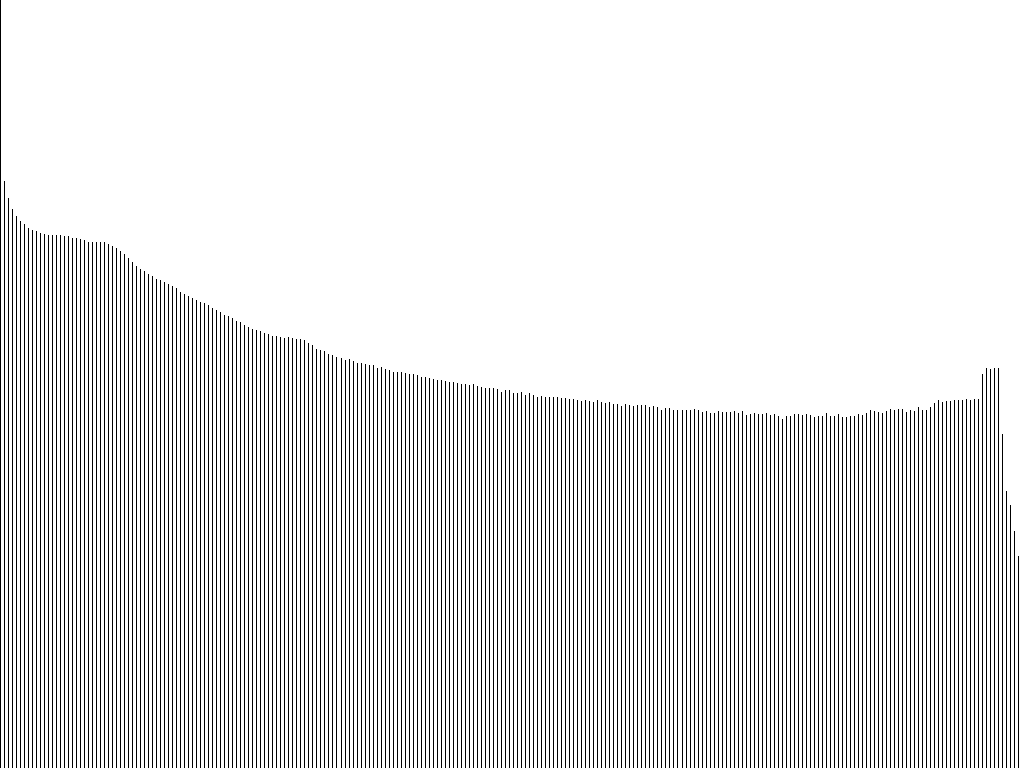}&
  \tikzpic{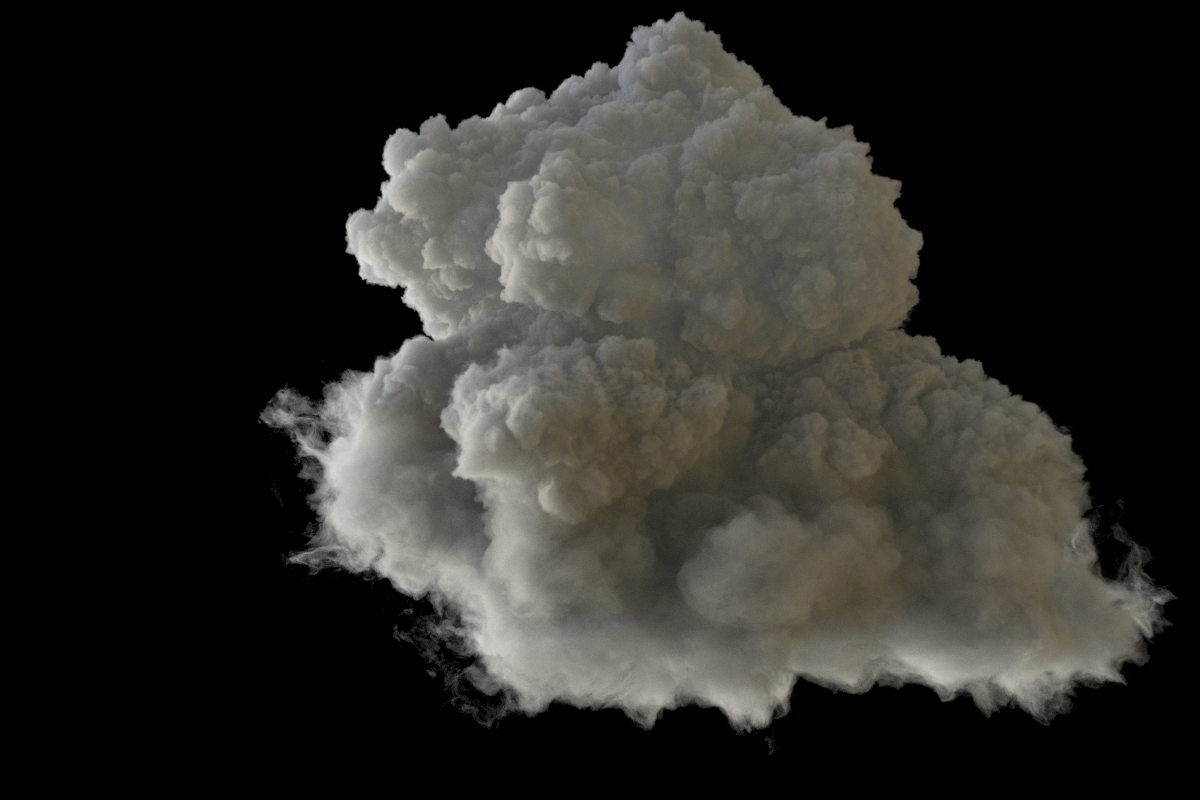}{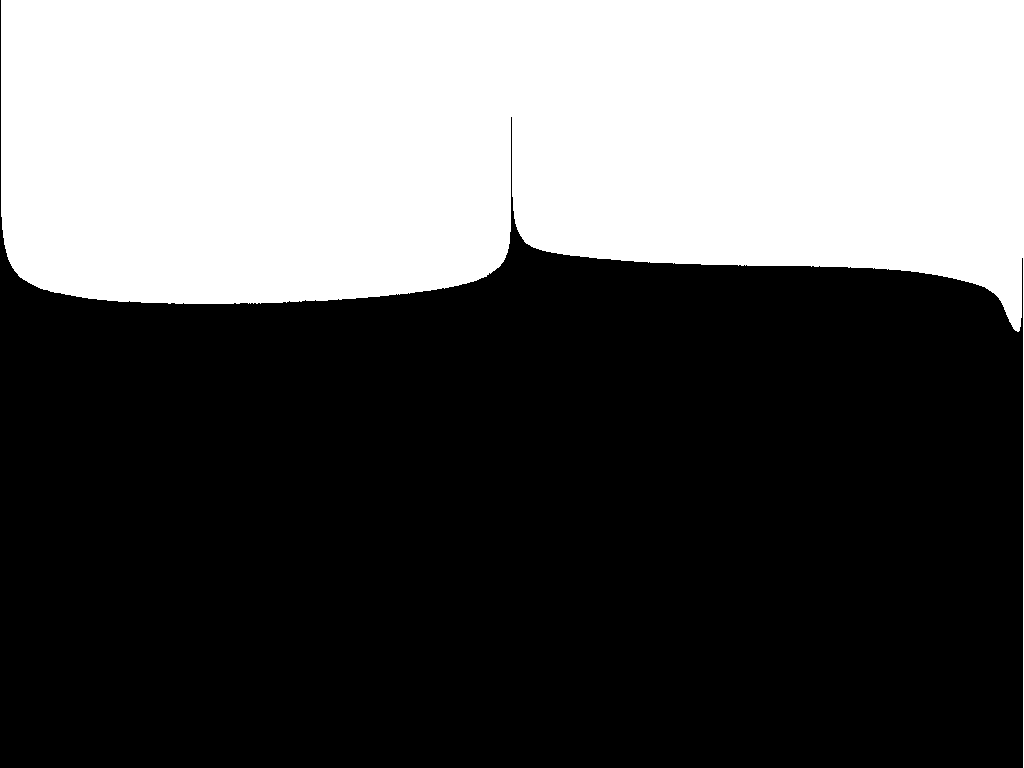}\\

  \small{Aneurism} &
  \small{Heptane} &
  \small{WDAS Cloud} \\
  \small{256 $\times$ 256 $\times$ 256} &
  \small{302 $\times$ 302 $\times$ 302} &
  \small{1987 $\times$ 1351 $\times$ 2449} \\

  \tikzpic{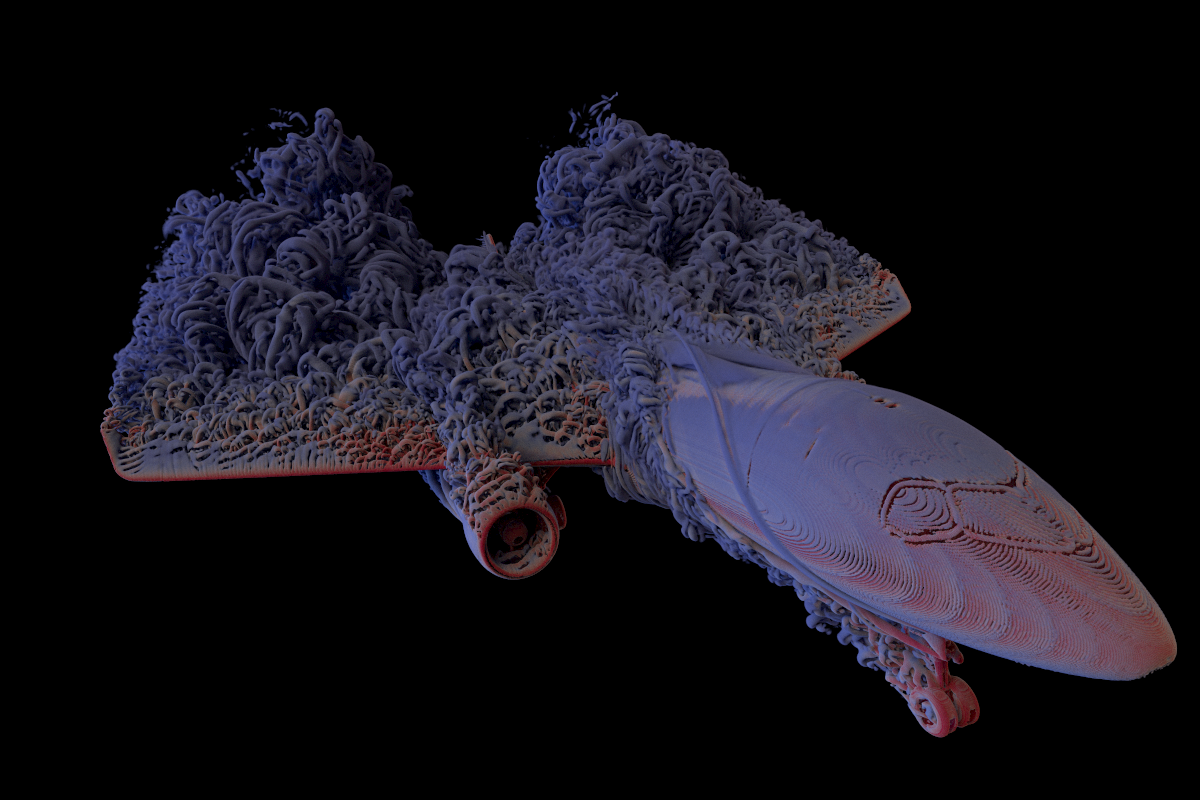}{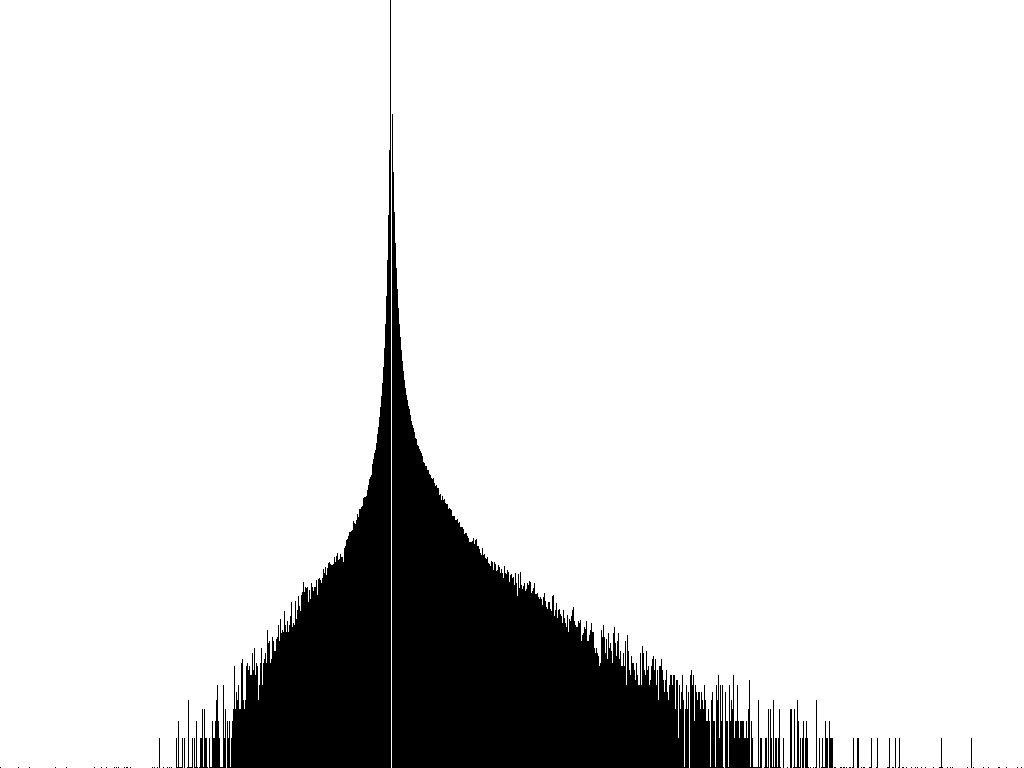}&
  \tikzpic{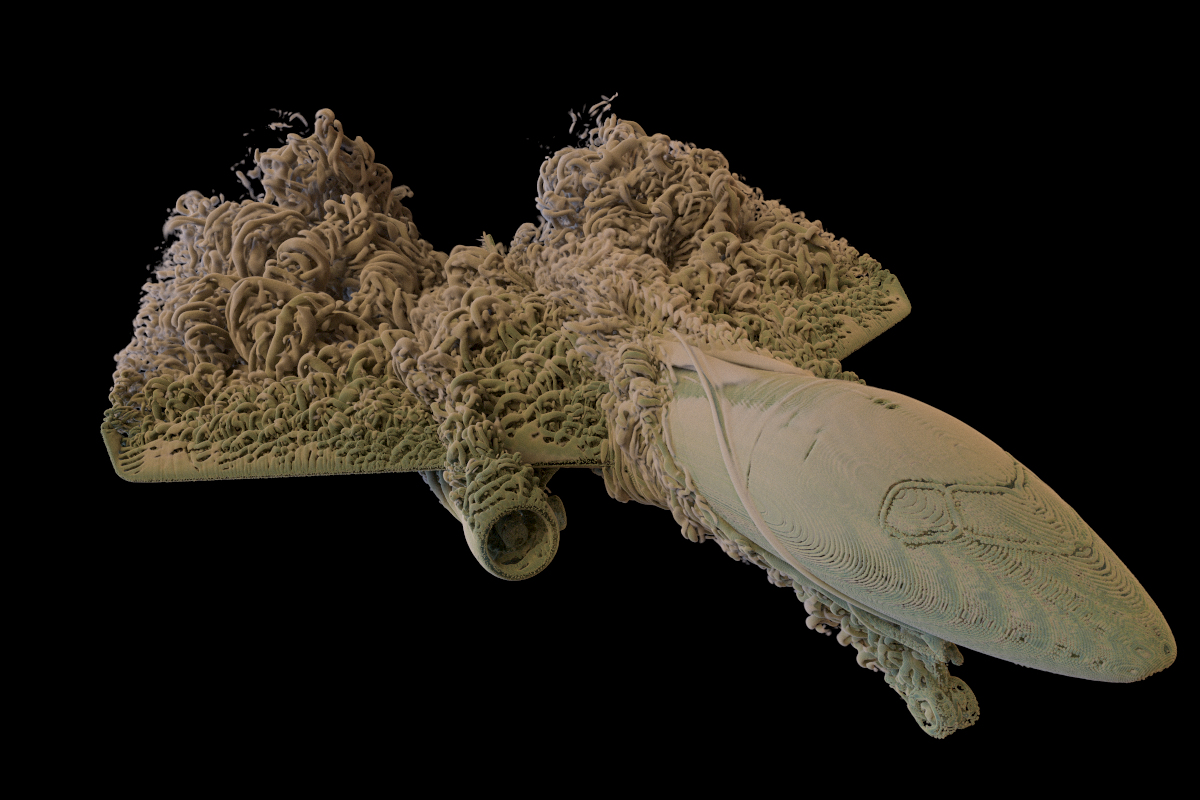}{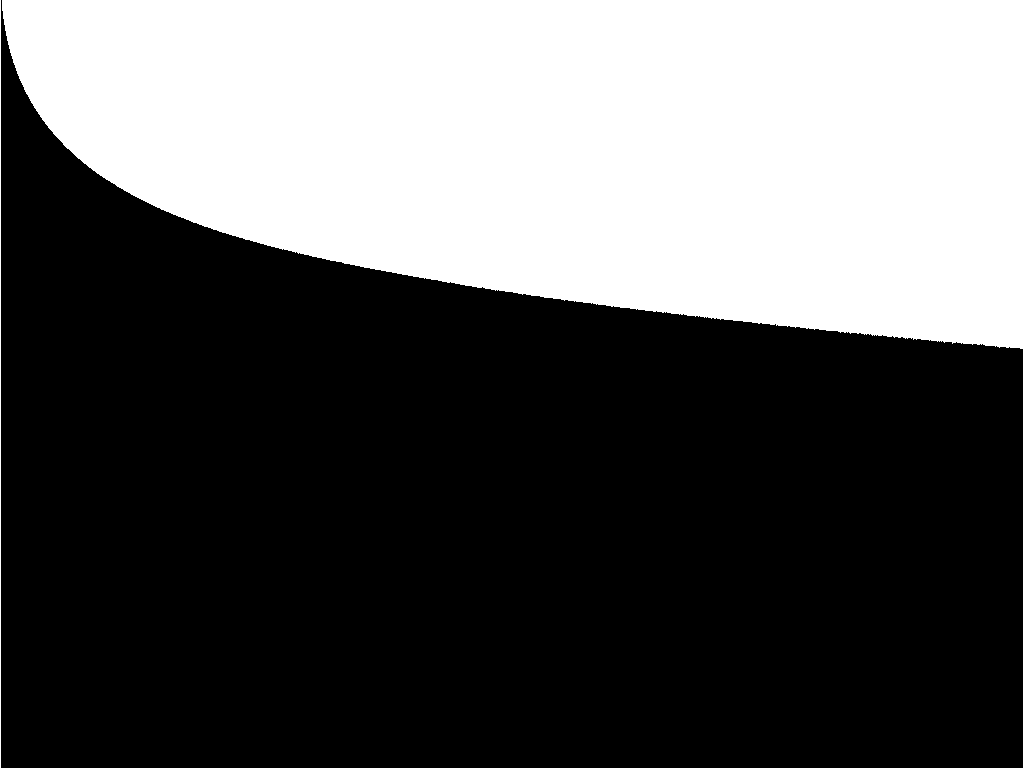}&
  \tikzpic{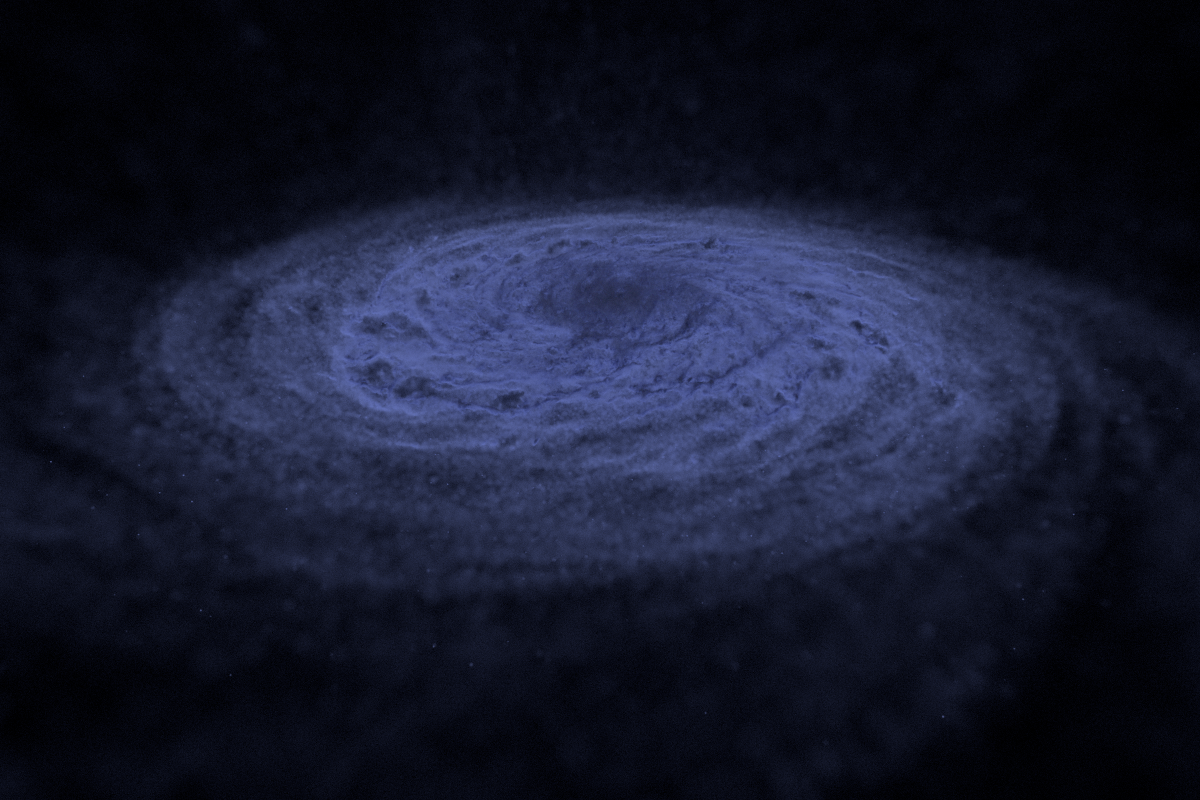}{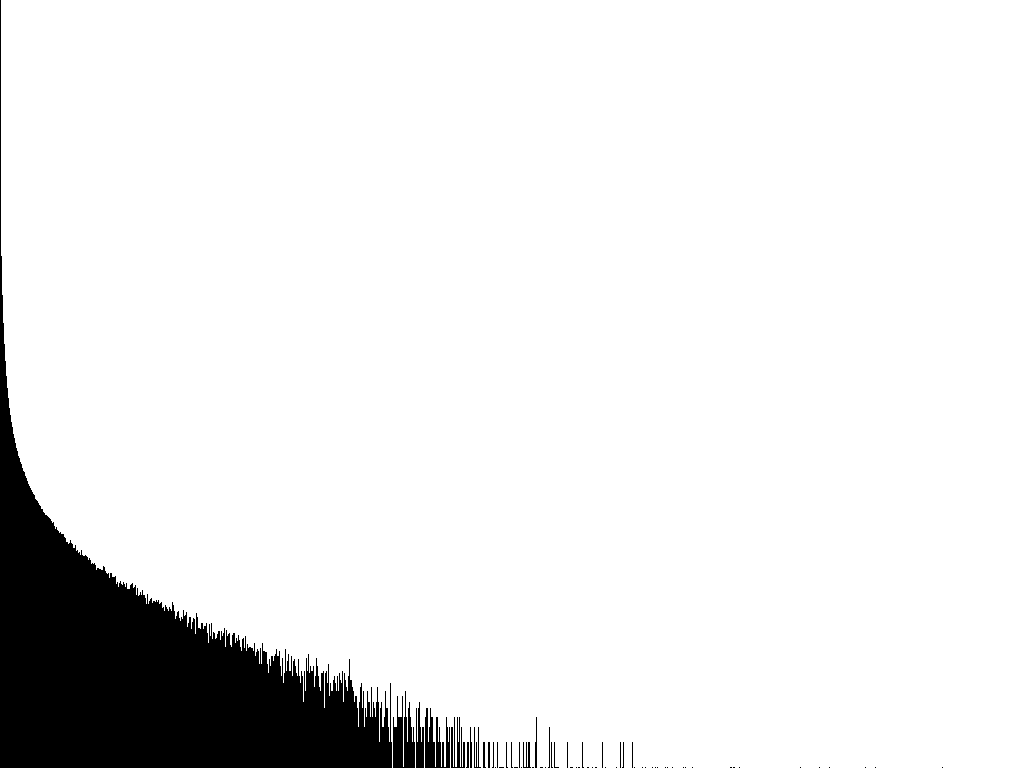}\\

  \small{Airplane Q-Vorticity} &
  \small{Airplane Q-Vorticity (filtered)} &
  \small{Galaxy B-field} \\
  \small{2450 $\times$ 1162 $\times$ 4000} &
  \small{2450 $\times$ 1160 $\times$ 4000} &
  \small{3000 $\times$ 3000 $\times$ 3000} \\

\end{tabular}
\end{center}
\vspace{-0.5em}
\caption{\label{tab:datasets}
Evaluation data sets. We use data sets of different sizes and with different
degrees of sparseness; in the bottom-left corner we show log-scale histograms
of the original structured-regular data.
}
\vspace{-2em}
\end{table*}
We evaluate the quality and performance of our compressed representation for
direct volume rendering. For that we use the structured-regular volume data
sets from \cref{tab:datasets}. The \code{aneurism} data set is particularly
sparse. The \code{WDAS cloud} data set by Walt Disney Animation
Studios~\cite{wdas-cloud} is originally available as a VDB and was resampled to a
structured-regular representation retaining the original 2K resolution.
As we also have access to the original VDB, this gives us the possibility to
compare back-and-forth conversion using our algorithm. The \code{airplane} data
set was simulated with OpenFOAM; we use the Q-criterion
vorticity field which is usually sparse compared to other variables.  The data
was voxelized using the MESIO library~\cite{MECA2022103100}, as detailed
in~\cite{Faltynkova2025}. We have access to two versions of that data set: one
that represents the vorticity, and a filtered version of that field where $Q
\in [0:500K]$, retaining only 4.4~\% of the original voxels. We also use a
high-resolution,
non-cosmological isolated galaxy data set for the evaluation. We utilize an example from
\cite{galaxy-dataset}, designed to resemble a Milky Way-type galaxy. The
simulation incorporates magnetohydrodynamics (MHD) along with standard subgrid
physics commonly employed in galaxy formation studies, including gas dynamics
with a magnetic field. The simulation was produces using ChaNga~\cite{changa}
and subsequently converted to a floating-point volume on a structured-regular grid.
To simplify the comparison we assume voxels to be represented with 32-bit
floating-point values; we convert \code{aneurism} and \code{heptane} to that
format whose voxels use 8-bit values.

\subsection{Compression Rate}
\begin{table}[th]
\begin{center}
  \setlength{\tabcolsep}{.6ex}
\begin{tabular}{r|c|c|c|c|c}
\toprule
Data Set     & Uncompressed & 1:8 & 1:4 & 1:2 & 1:1 \\
\midrule

Aneurism     & 67.1~MB & 9.3~MB & 16~MB  & 16~MB  & 16~MB \\
Heptane      & 110~MB & 15~MB  & 30~MB  & 35~MB  & 42~MB \\
WDAS         & 26.3~GB & 3.5~GB & 5.8~GB & 5.9~GB & 6.2~GB \\
Airplane     & 45.6~GB & 5.7~GB & 12~GB  & 24~GB  & 48~GB\\
Airplane (f) & 45.5~GB & 4.5~GB & 4.5~GB & 4.5~GB & 4.6~GB\\
Galaxy       & 108~GB  & 14~GB  & 29~GB  & 57~GB  & 91~GB\\

\bottomrule
\end{tabular}
\end{center}
\caption{\label{tab:rates}%
Actual size in memory of the NanoVDB representation when compressing at
different rates.
}
\end{table}
We first evaluate if, by using our fixed-rate encoding, the target compression
rate is achieved and compress the data set with different ratios from 1:8 to
1:1 (the latter meaning the targeted reduction in size is zero percent). In the
latter case the algorithm is guaranteed to be lossless but will only activate
voxels that are not considered empty, so the actual size will be lower than the
original one. To assess this we report the actual size in memory of the resulting
NanoVDBs in \cref{tab:rates}.
\begin{table}[th]
\begin{center}
  \setlength{\tabcolsep}{.3ex}
\begin{tabular}{r|c|c|c|c|c|c|c}
\toprule
        &
\includegraphics[width=0.11\columnwidth]{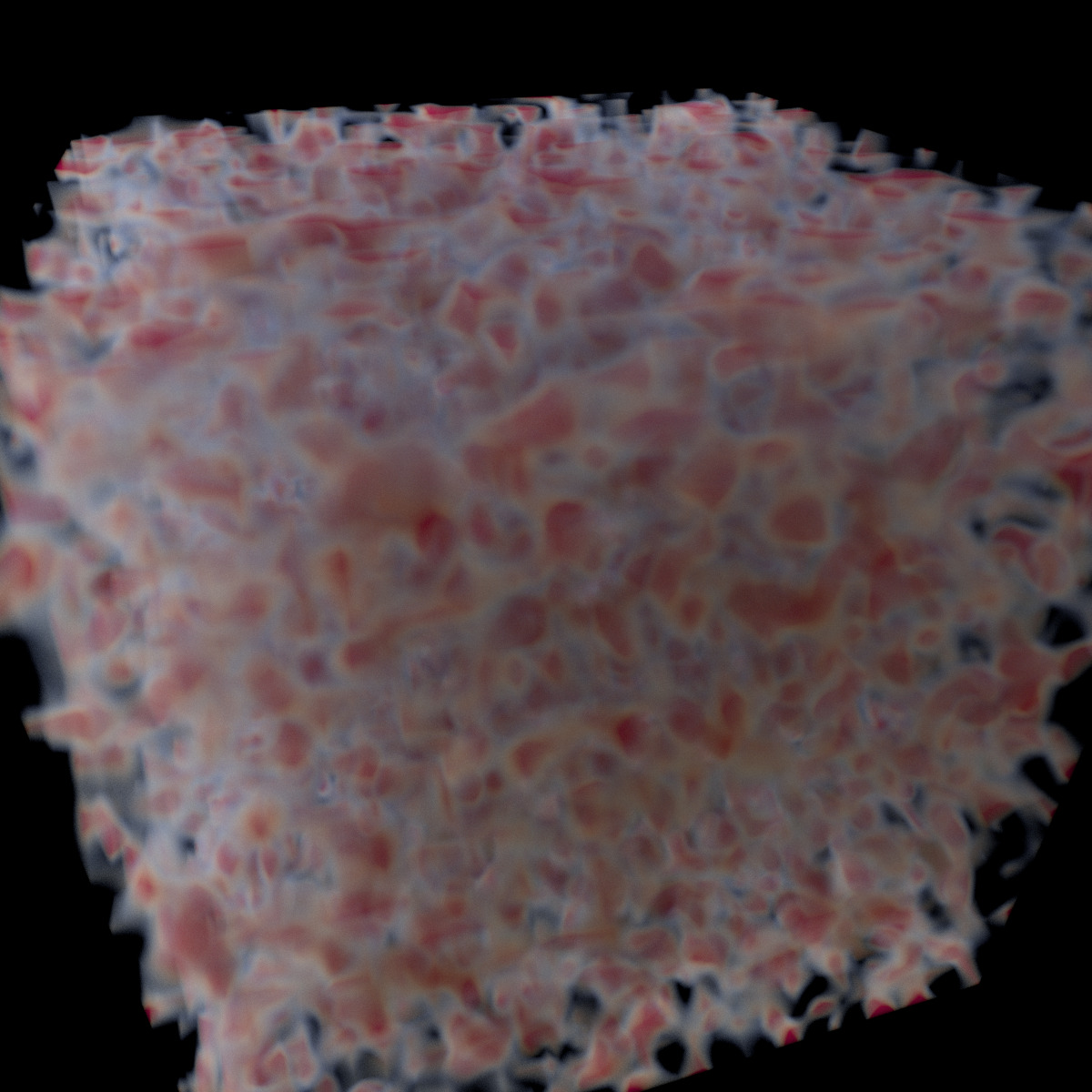} &
\includegraphics[width=0.11\columnwidth]{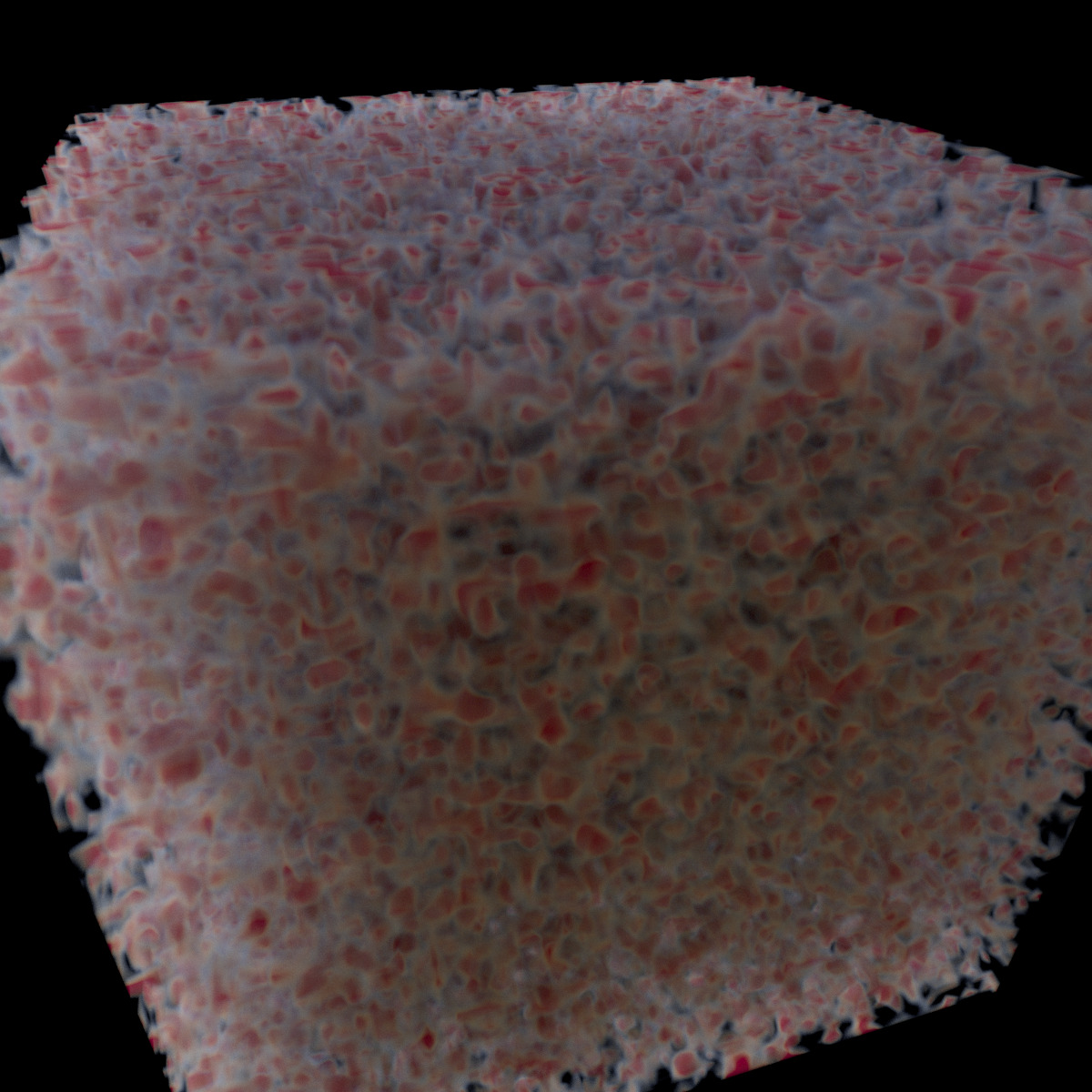} &
\includegraphics[width=0.11\columnwidth]{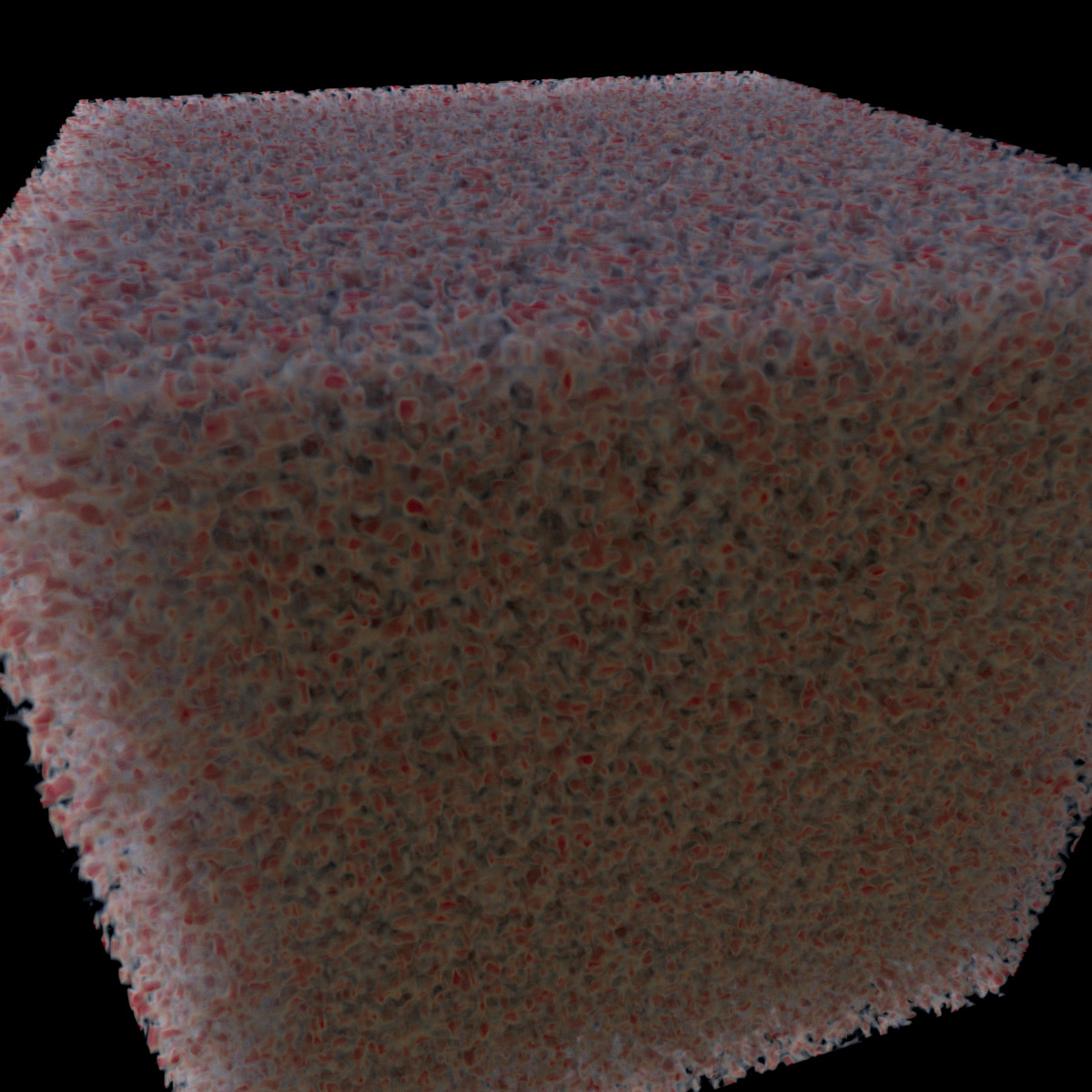} &
\includegraphics[width=0.11\columnwidth]{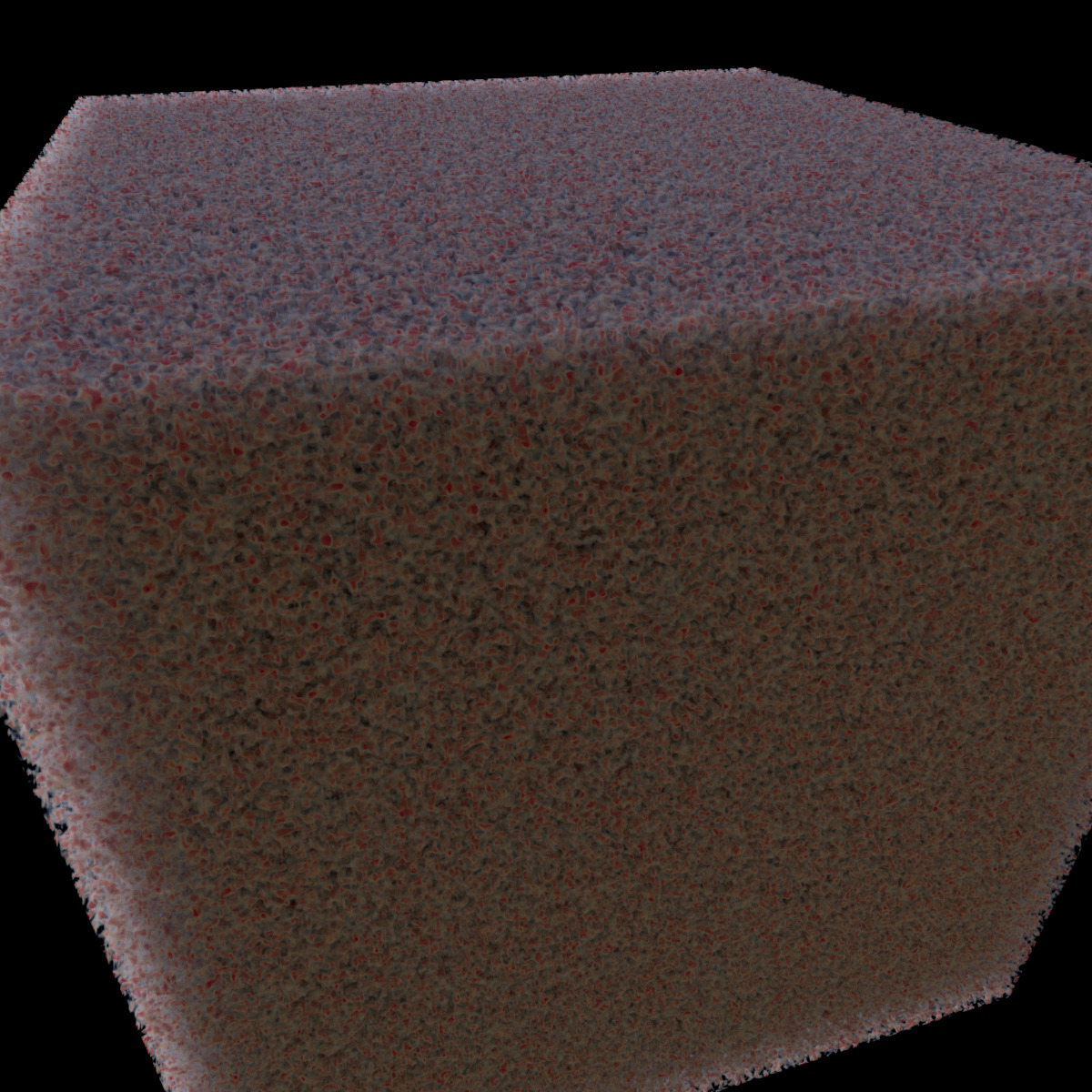} &
\includegraphics[width=0.11\columnwidth]{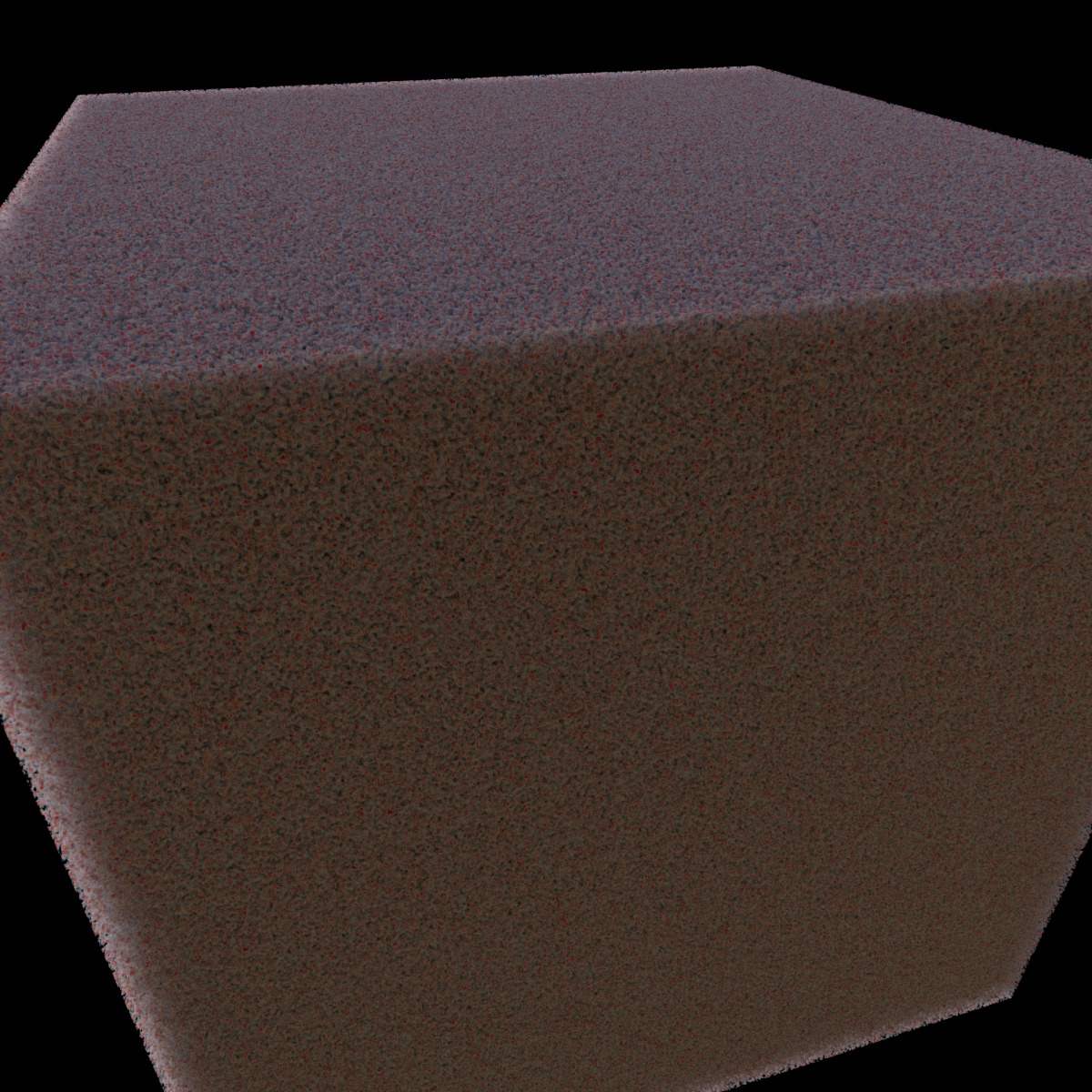} &
\includegraphics[width=0.11\columnwidth]{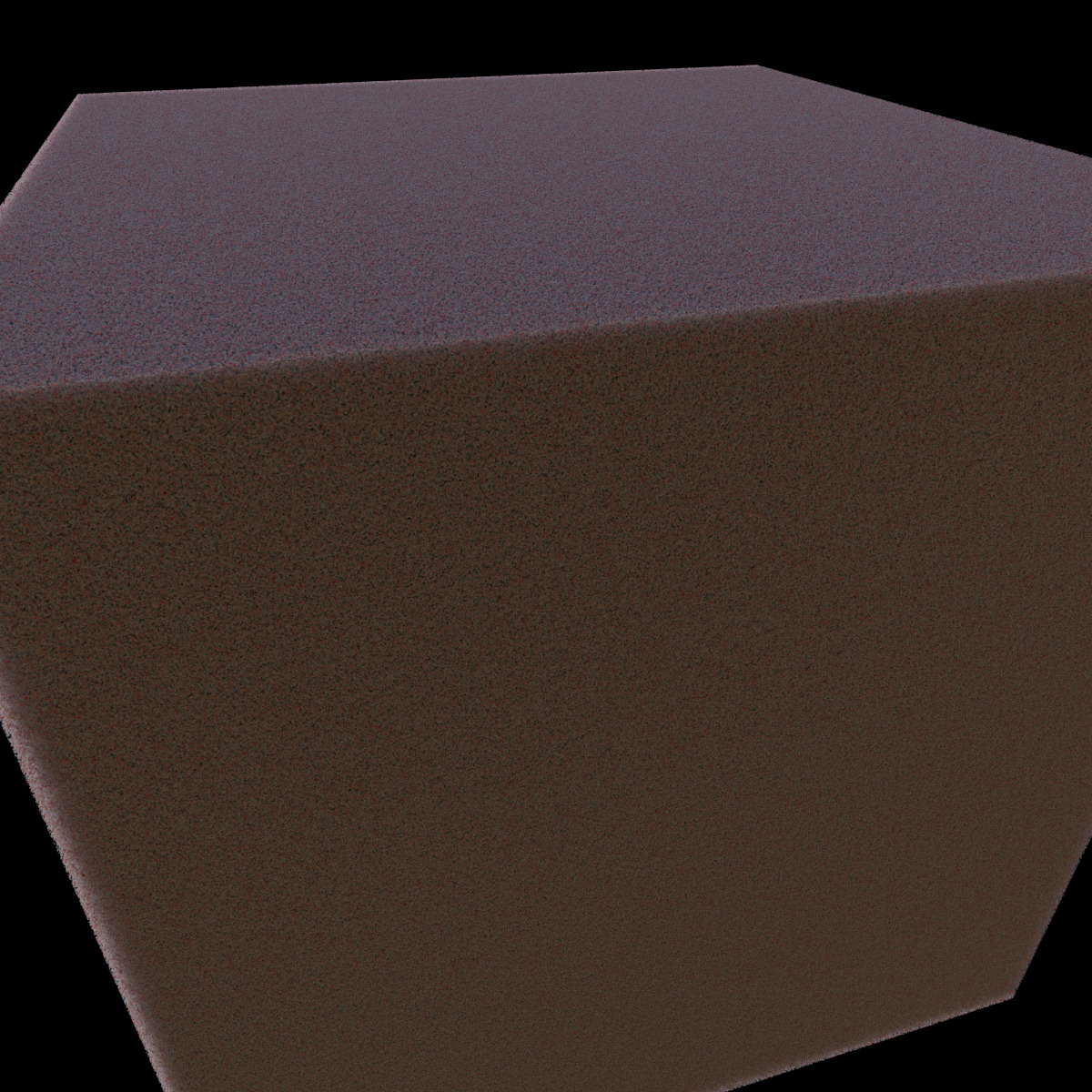} &
\includegraphics[width=0.11\columnwidth]{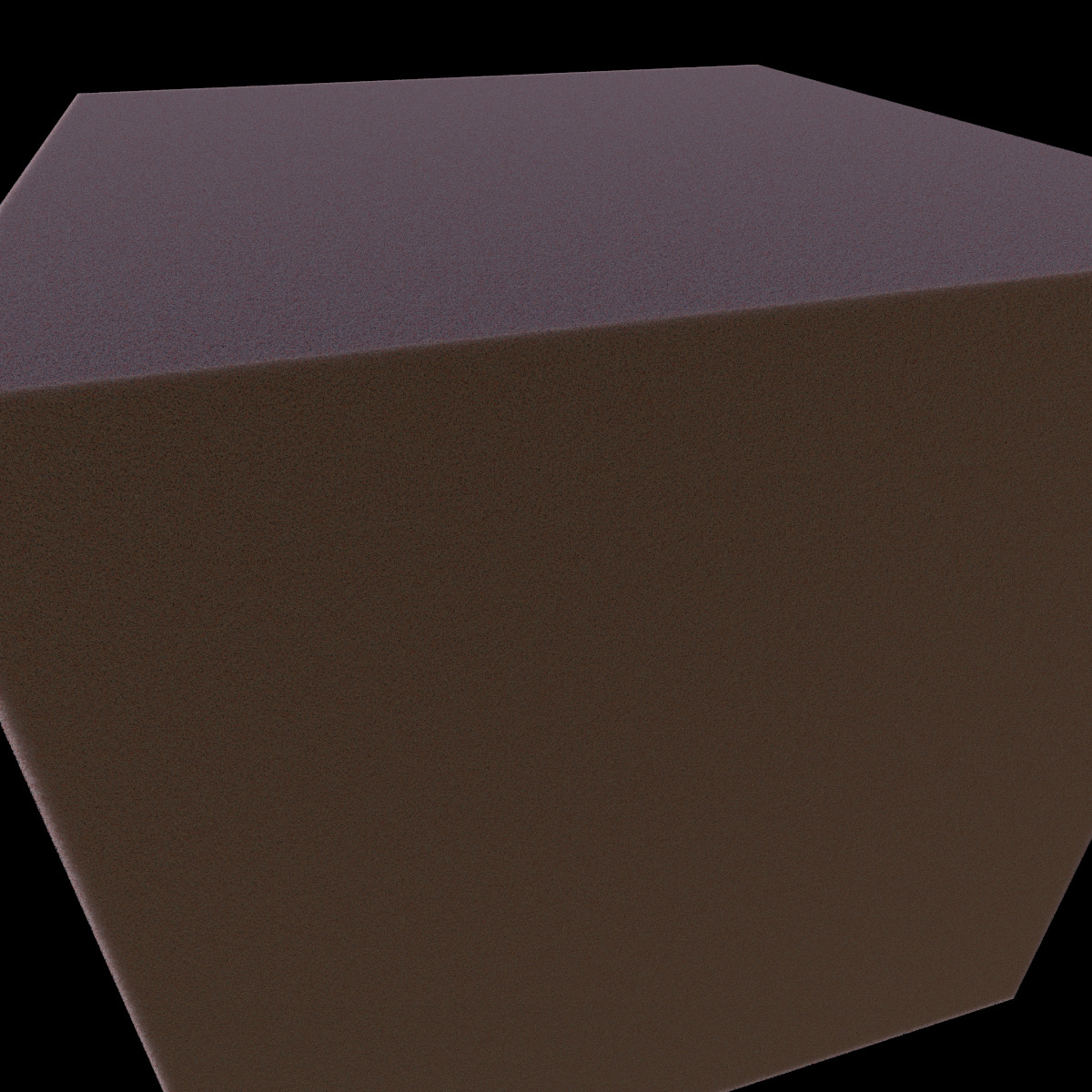} \\
\toprule
        & $32^3$ & $64^3$ & $128^3$ & $256^3$ & $512^3$ & $1024^3$ & $2048^3$ \\
\midrule
\textbf{Size} & \\
CUDA    & 131~KB & 1.0~MB & 8.4~MB  & 67~MB   & 537~MB  & 4.3~GB   & 34~GB    \\
VDB     & 440~KB & 1.4~MB & 9.1~MB  & 71~MB   & 564~MB  & 4.5~GB   & 36~GB    \\
\midrule
\textbf{FPS} & \\
CUDA&	42.0&	27.0&	22.1&	19.9&	18.4&	17.1&	16.3 \\
VDB &	31.9&	22.0&	17.8&	16.2&	15.2&	14.0&	13.3 \\

\bottomrule
\end{tabular}
\end{center}
\caption{\label{tab:worstcase}%
\rev{Size in memory and frame rate per second (FPS) when compressing truly dense
data from a noise texture. We compare ours (VDB) against a renderer using dense
GPU textures (CUDA).} Even in this adversarial case the memory overhead for
large-scale data sets does not exceed 10~\%.
\vspace{-1em}
}
\end{table}

We are also interested in the worst case we have to expect when losslessly
compressing a volume that is not sparse. This is an extreme case that we
simulate by generating synthetic volume data sets. We generate noise textures
of different sizes to fill the structured-regular grids with voxels so that no
local neighborhood is considered empty by our algorithm. We report results for
this experiment in \cref{tab:worstcase}. For large-scale data we observe that
the overhead does not exceed 10~\% of the original size. \rev{We also found,
although without a formal evaluation, that there is no correlation between
sparseness and render performance and that NanoVDB performs equally well on
data that is dense.} This suggests that VDBs are a viable replacement even for
dense textures.

\subsection{Compression Quality}
\begin{table*}[th]
\begin{center}
\setlength{\tabcolsep}{1pt}
\begin{tabular}{r|cc|cc|cc|cc|cc|cc|cc|cc|cc|cc}
\toprule
    Rate & \multicolumn{2}{c|}{10~\%} &
           \multicolumn{2}{c|}{20~\%} &
           \multicolumn{2}{c|}{30~\%} &
           \multicolumn{2}{c|}{40~\%} & 
           \multicolumn{2}{c|}{50~\%} & 
           \multicolumn{2}{c|}{60~\%} & 
           \multicolumn{2}{c|}{70~\%} & 
           \multicolumn{2}{c|}{80~\%} & 
           \multicolumn{2}{c|}{90~\%} & 
           \multicolumn{2}{c}{100~\% (1:1)} \\
Data Set & MSE & PSNR & MSE & PSNR & MSE & PSNR & MSE & PSNR & MSE & PSNR & MSE & PSNR & MSE & PSNR & MSE & PSNR & MSE & PSNR & MSE & PSNR \\
\midrule
Aneur.\    & 1e-6 & 59.0 & 2e-9 & 86.7 & 0.00 & \INF & 0.00 & \INF & 0.00 & \INF & 0.00 & \INF & 0.00 & \INF & 0.00 & \INF & 0.00 & \INF & 0.00 & \INF \\
Heptane     & 5e-4 & 32.7 & 1e-5 & 49.0 & 0.00 & \INF & 0.00 & \INF & 0.00 & \INF & 0.00 & \INF & 0.00 & \INF & 0.00 & \INF & 0.00 & \INF & 0.00 & \INF \\
WDAS        & 4e-3 & 23.9 & 6e-9 & 81.9 & 0.00 & \INF & 0.00 & \INF & 0.00 & \INF & 0.00 & \INF & 0.00 & \INF & 0.00 & \INF & 0.00 & \INF & 0.00 & \INF \\
Airplane    & 7e-9 & 81.8 & 6e-9 & 82.3 & 5e-9 & 82.9 & 4e-9 & 83.6 & 4e-9 & 84.4 & 3e-9 & 85.4 & 2e-9 & 86.7 & 1e-9 & 88.5 & 7e-10& 91.7 & 1e-10& 119  \\
Airpl.\ (f)& 0.00 & \INF & 0.00 & \INF & 0.00 & \INF & 0.00 & \INF & 0.00 & \INF & 0.00 & \INF & 0.00 & \INF & 0.00 & \INF & 0.00 & \INF & 0.00 & \INF \\
Galaxy      & 2e-16& 156  & 5e-17& 163  & 2e-18& 169  & 6e-19& 175  & 7e-19& 182  & 7e-20& 192  & 1e-21& 209  & 0.00 & \INF & 0.00 & \INF & 0.00 & \INF \\
\bottomrule
\end{tabular}
\end{center}
\vspace{-0.5em}
\caption{\label{tab:stats}
Compression rate statistics. We report mean squared error (MSE, lower is
better) and peak signal-to-noise ratio (PSNR, in dB, higher is better) for the
data sets from \cref{tab:datasets}. PSNR $ = \infty$ indicates lossless
compression.
}
\vspace{-2em}
\end{table*}
We first evaluate the compression quality using similarity metrics comparing
the compressed representation with their dense texture counterparts.
These results are summarized in \cref{tab:stats}; we tabulate mean squared
error (MSE) and peak signal-to-noise-ratio (PSNR) for the test data sets.
As expected, the measured quality is a function of sparseness.
\begin{figure}[tb]
\centering
  \includegraphics[width=0.48\columnwidth]{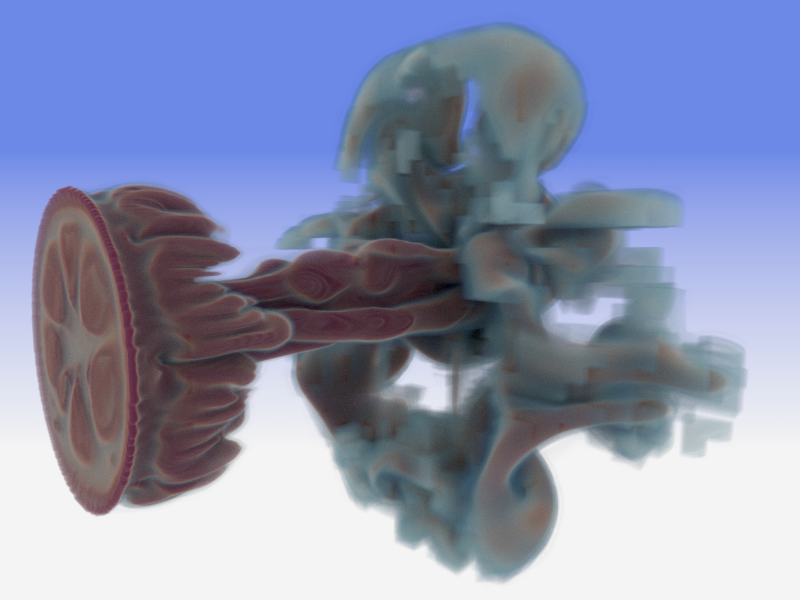}
  \includegraphics[width=0.48\columnwidth]{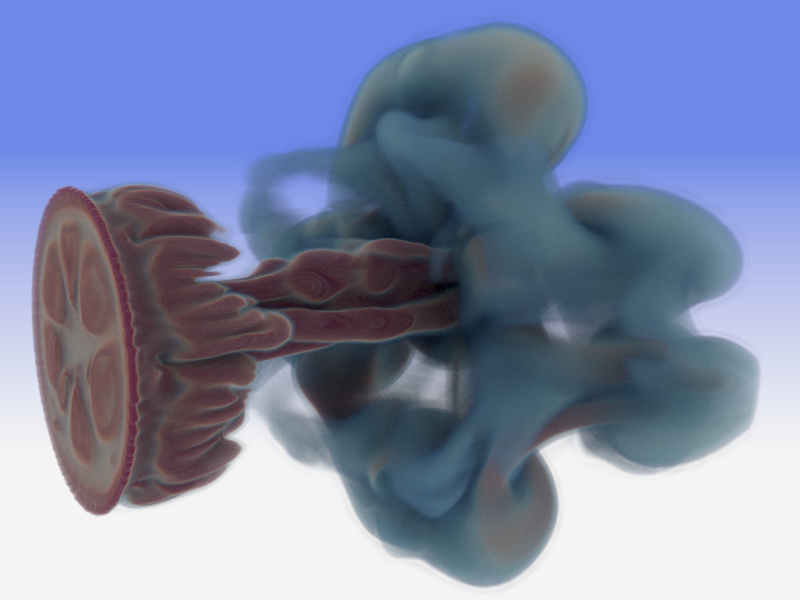}\\
  \includegraphics[width=0.48\columnwidth]{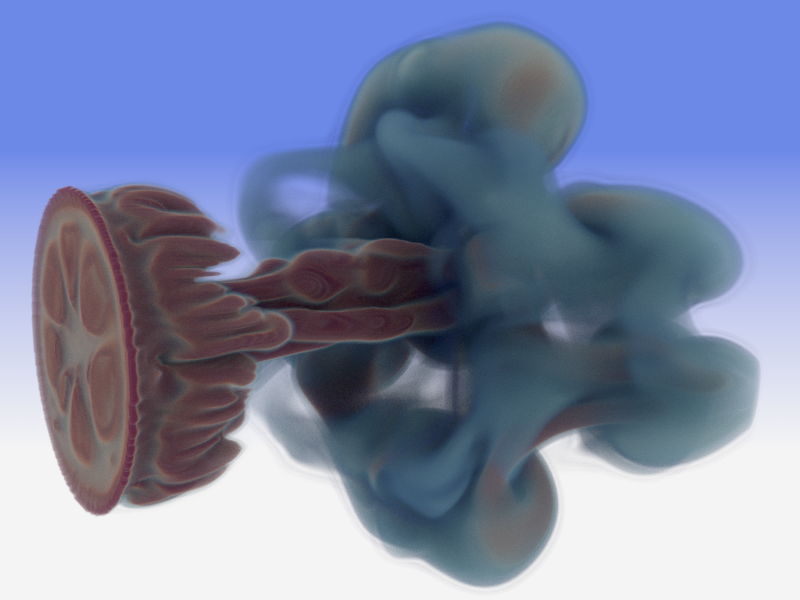}
  \includegraphics[width=0.48\columnwidth]{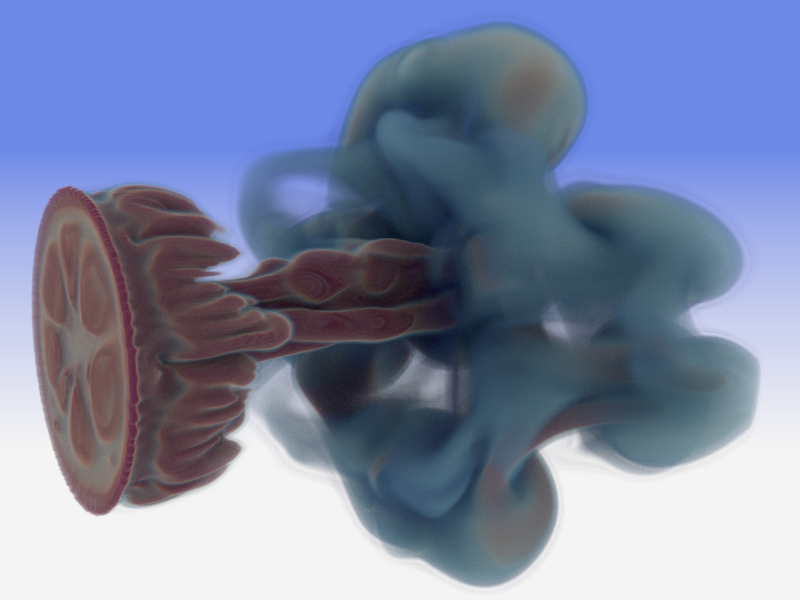}
  \vspace{-1em}
  \caption{\label{fig:heptane}
  Our algorithm assigns homogeneous blocks to regions that are
  considered nearly empty, and voxelizes regions that are not.
  The compression rates shown are: Top-left: 1:8, top-right: 1:5, bottom-left:
  1:4, bottom-right: 1:3. Whether blocks become visible depends on
  the compression rate and the transfer function.
  }
\vspace{-2em}
\end{figure}
While giving a first impression, similarity metrics traditionally have
limitations making it hard to draw conclusions relating the metric to the
perceived quality. What distinguishes our compression method from other methods
is that we encode some values---those that are closer to the VDB background
value; i.e., some local features will be perceived as blocky or have missing
features, while other volumetric regions will appear as if compressed with a
lossless algorithm (cf. \cref{fig:heptane}). This is particularly hard to assess
from just looking at similarity metrics like MSE or PSNR. What we can conclude
from those results though is at which compression rate our algorithm achieves
lossless encoding. We observe that many of our data sets can be compressed by
25~\% and higher without any quality loss at all.
\begin{figure}[tb]
\centering
  \includegraphics[width=0.48\columnwidth]{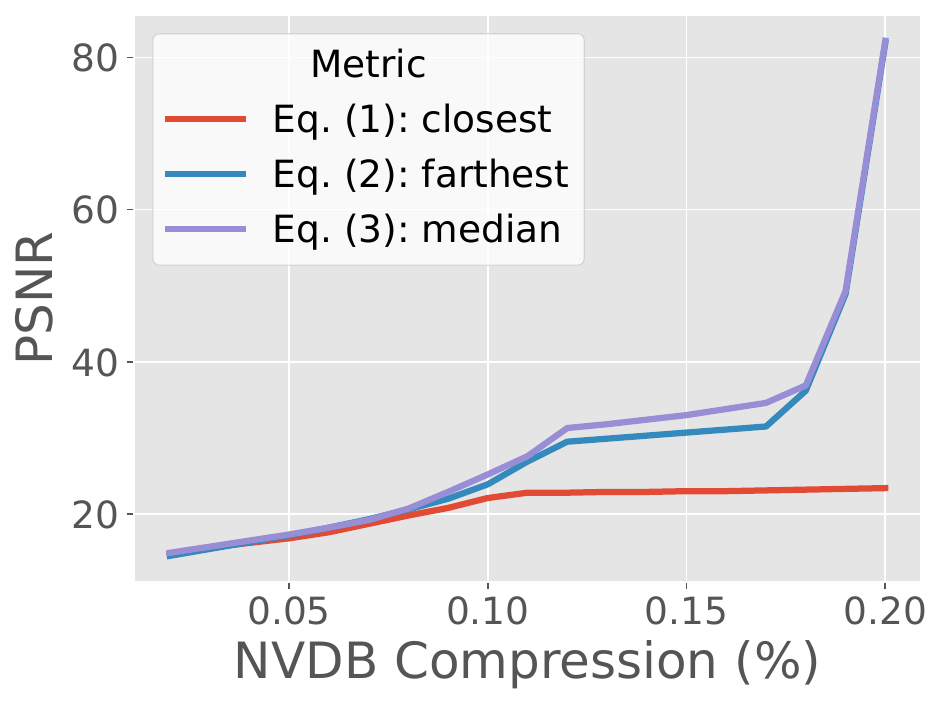}
  \vspace{-1em}
  \caption{\label{fig:distance}
  Similarity metrics and their impact on quality. The results represent PSNR
  for different compression rates using the WDAS cloud data set. Graphs for the
  other data sets look similar; we found the farthest and median point-in-range
  similarity metrics to outperform the closest point-in-range metric.
  \vspace{-2em}
  }
\end{figure}

We then evaluate the impact of the closest, farthest, and median point-in-range
metrics from \cref{sec:algo}. We exemplarily present results for the WDAS cloud
in \cref{fig:distance}, which indicate superiority of the farthest and median
over the closest point-in-range metric. We observed similar outcomes for
experiments with the data sets not shown here, leading us to the conclusion
that these two metrics seem generally superior.

\subsection{Comparison with Dense Texture Multi-GPU Volume Renderer}
\begin{table}[th]
\begin{center}
  \setlength{\tabcolsep}{.6ex}
\begin{tabular}{r|c|c|c}
\toprule
Data Set     & \# GPUs (dense) & CUDA texture & VDB (ours) \\
\midrule
Aneurism     & 1 & 75.9 & 55.8 \\
Heptane      & 1 & 49.2 & 42.1 \\
WDAS         & 1 & 25.4 & 24.2 \\
Airplane     & 2 & 20.0 & 18.2 \\
Airplane (f) & 2 & 19.5 & 17.4 \\
Galaxy       & 3 & 18.6 & 34.8 \\
\bottomrule
\end{tabular}
\end{center}
\caption{\label{tab:framerates}
Frame rates achieved for volumetric path tracing. We compare our VDB encoding
to encoding the structured data with dense 3D CUDA textures. In that case,
multi-GPU rendering is required for the airplane and galaxy data sets, where we
also report the number of GPUs used. 
}
\end{table}
We compare the rendering performance of VDBs to that of using random access
with dense textures. For that we use the Barney renderer with the extensions
described in \cref{sec:renderer} and either activate or deactivate VDB
compression. In the former case, Barney will encode the volume data sets using
VDB as described above, while in the latter case the structured-regular grid
representation is used. Barney represents structured volumes with 3D CUDA
textures.

We use an NVIDIA A100 multi-GPU node for the evaluation. Not all our data
sets fit into the 40~GB GPU memory, so we have to use two GPUs for the aircraft
and three GPUs for the galaxy data sets, respectively. Barney uses data parallel
rendering in this case, distributing the data evenly across the available GPUs,
and uses ray queue cycling~\cite{rqs} for volume path tracing.

We present average frame
rates with our volume path tracer for the rendered images in \cref{tab:framerates}.
These frame rate estimates indicate how long it takes to render noisy
convergence frames; the images in \cref{tab:datasets} represent converged
results over 1K convergence frames. To estimate the framerates we chose the $1:8$
compression ratio across all data sets, but note that framerates are stable
within small error margins regardless of the size of the VDB. While not faster,
we observe that software trilinear interpolation with NanoVDB achieves
competitive frame rates compared to using 3D CUDA textures. As the NanoVDBs
always fit into one GPU, our algorithm outperforms dense textures on the galaxy
data set that requires rendering on three GPUs.

\subsection{\rev{Quality} Comparison with ZFP}
\begin{figure}[tb]
\centering
  \includegraphics[width=0.48\columnwidth]{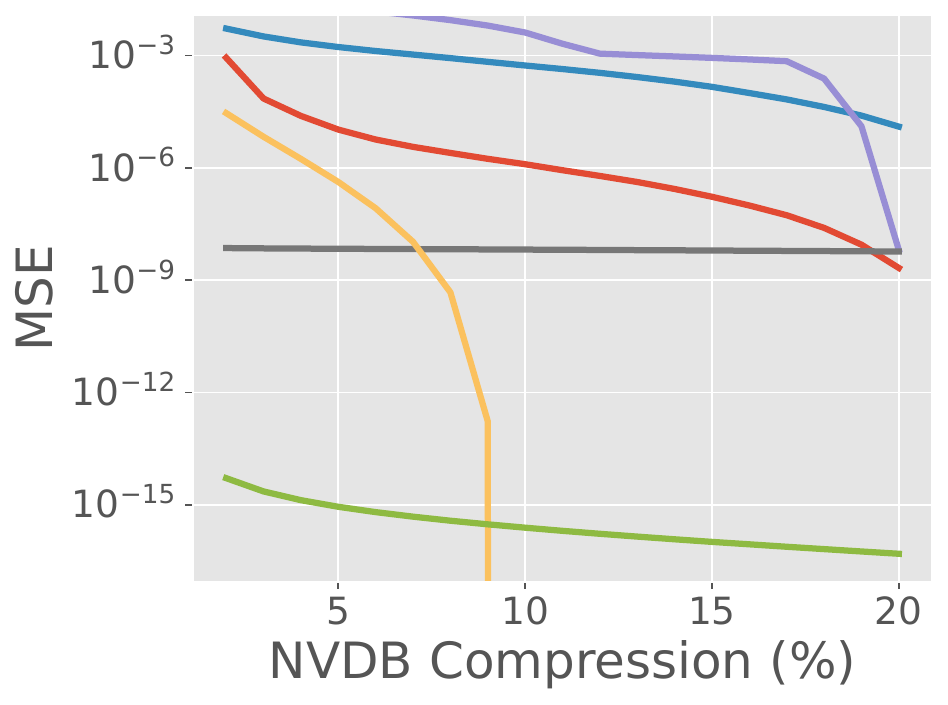}
  \includegraphics[width=0.48\columnwidth]{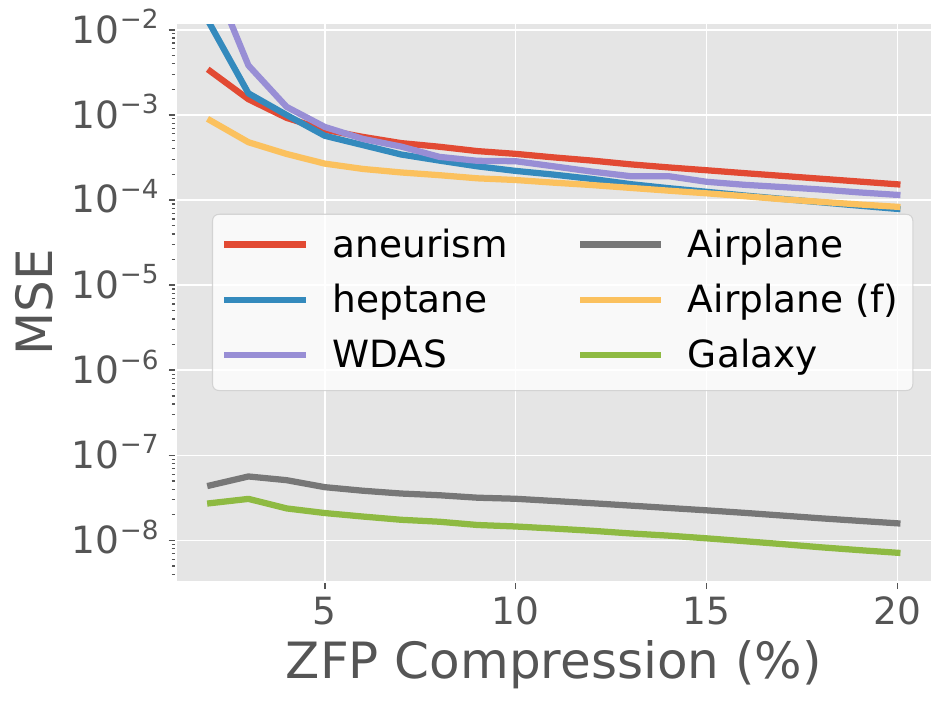}\\
  \includegraphics[width=0.48\columnwidth]{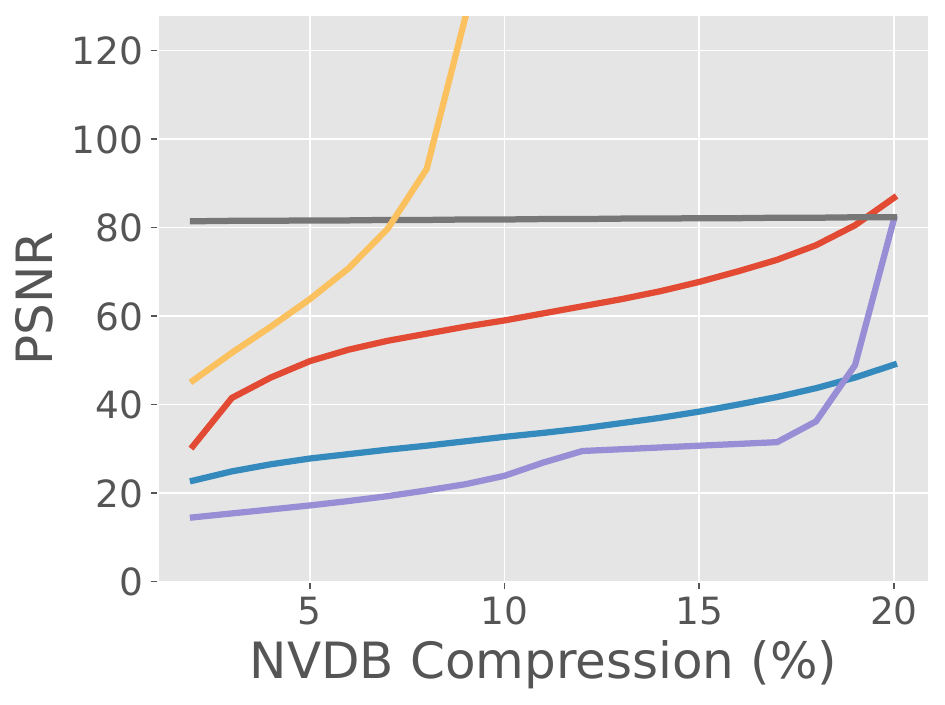}
  \includegraphics[width=0.48\columnwidth]{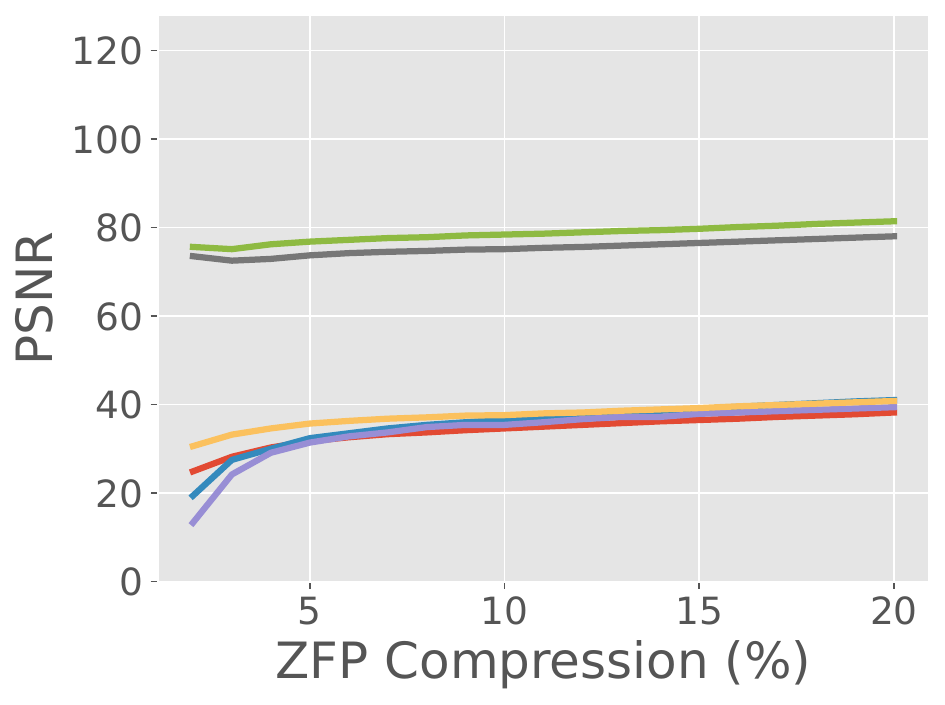}
  \vspace{-1em}
  \caption{\label{fig:zfp-comparison}
  Quality comparison of our method with ZFP. We report mean squared error (MSE,
  lower is better) and peak signal-to-noise ratio (PSNR, in dB, higher is
  better) for the data sets from \cref{tab:datasets}, for compression rates
  between 1-20~\%.
  \vspace{-2em}
  }
\end{figure}
\begin{figure*}[tb]
\centering
  \teasertikz{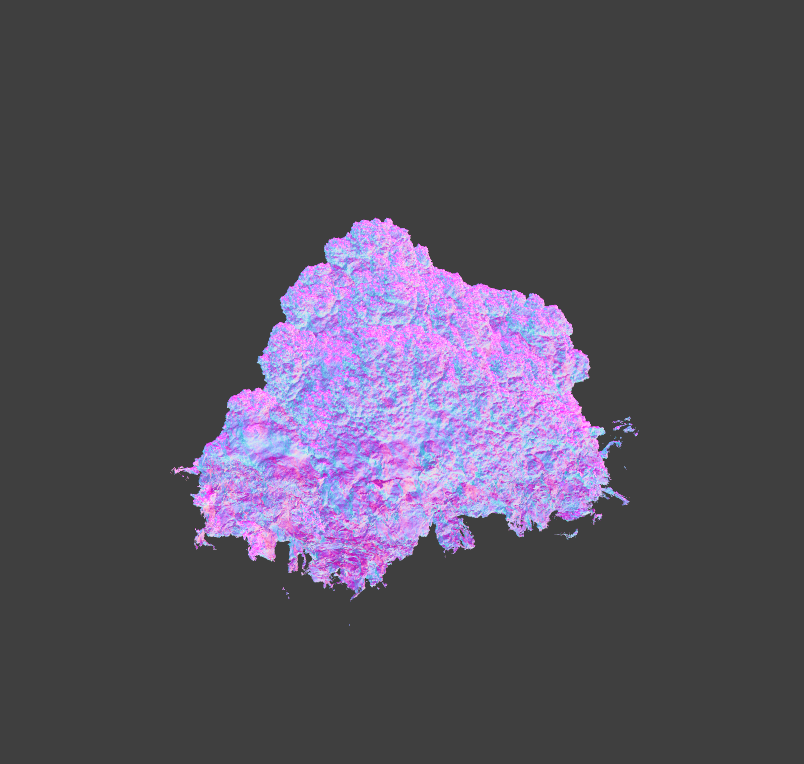}{reference}
  \teasertikz{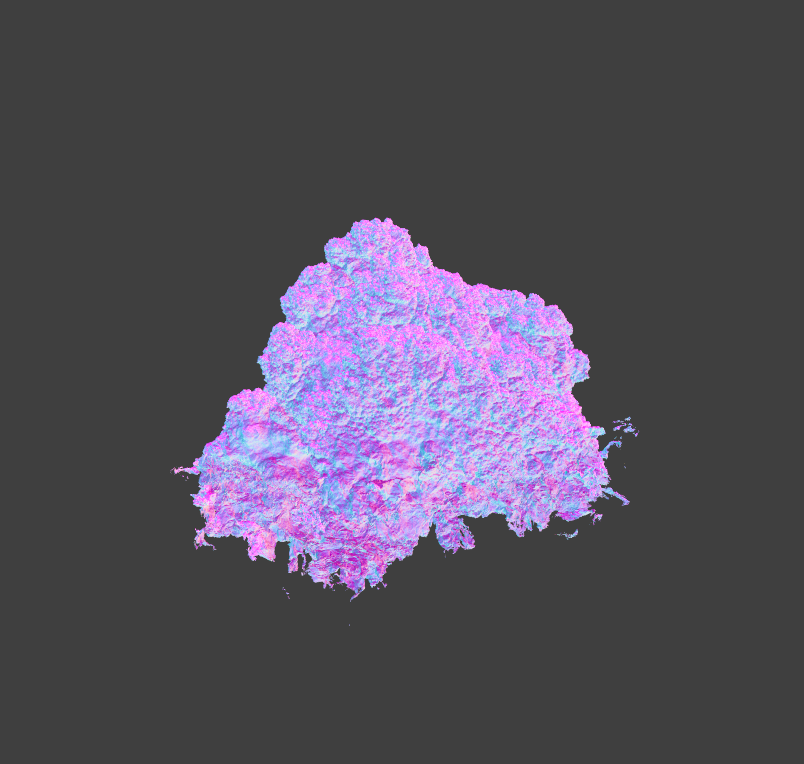}{1:10 (VDB)}
  \teasertikz{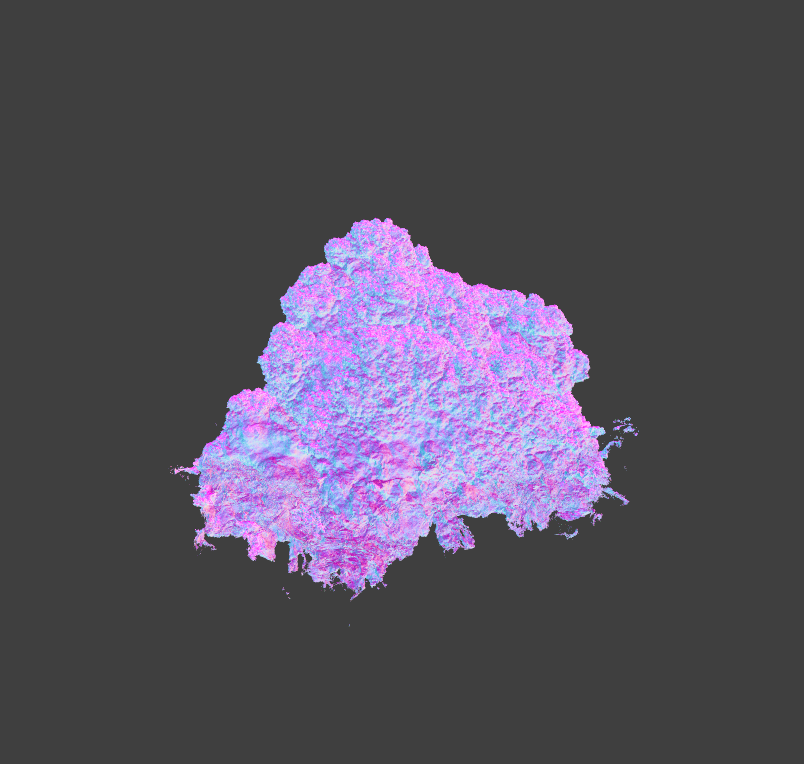}{1:10 (ZFP)}
  \teasertikz{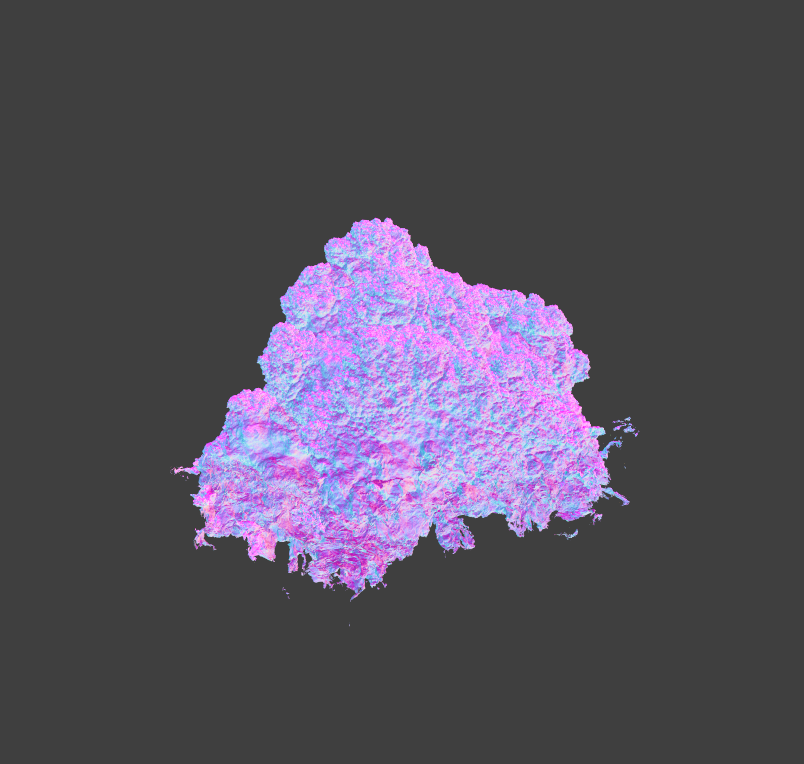}{1:6 (VDB)}
  \teasertikz{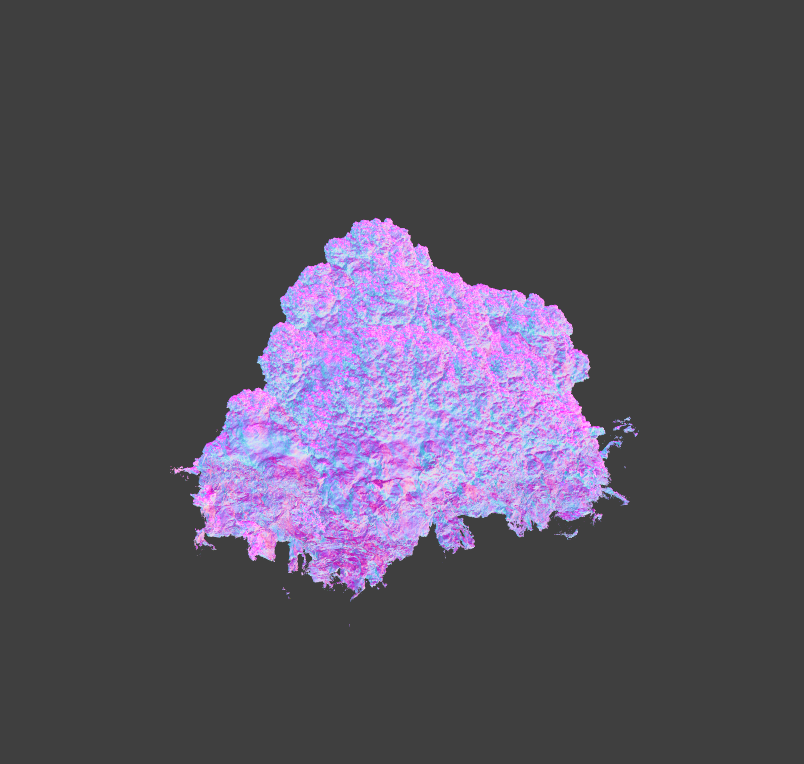}{1:6 (ZFP)}
  \vspace{-10pt}

  \teasertikz{png/white-pixel.jpg}{}
  \teasertikz{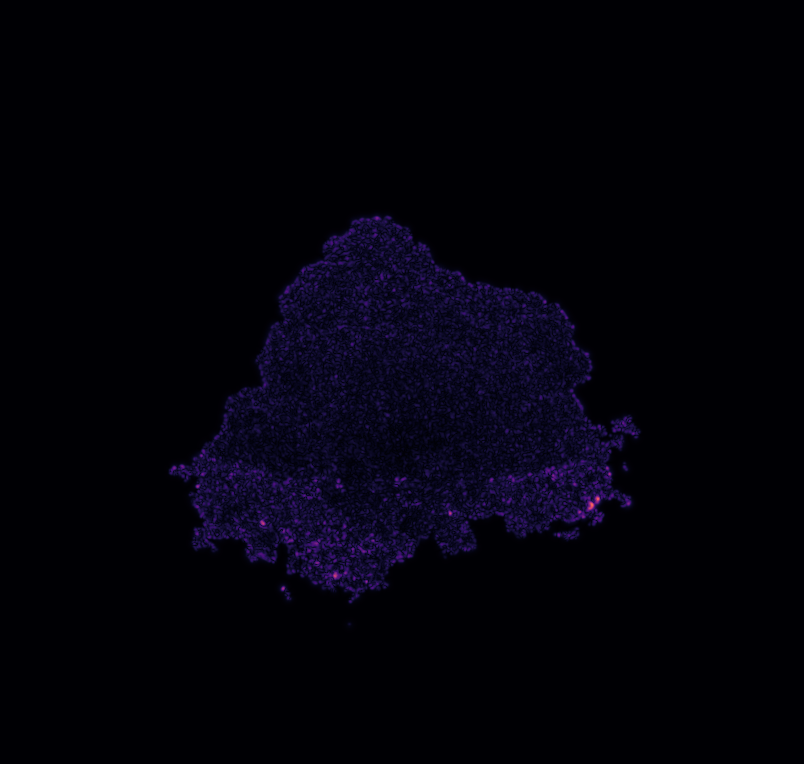}{\FLIP: 0.013}
  \teasertikz{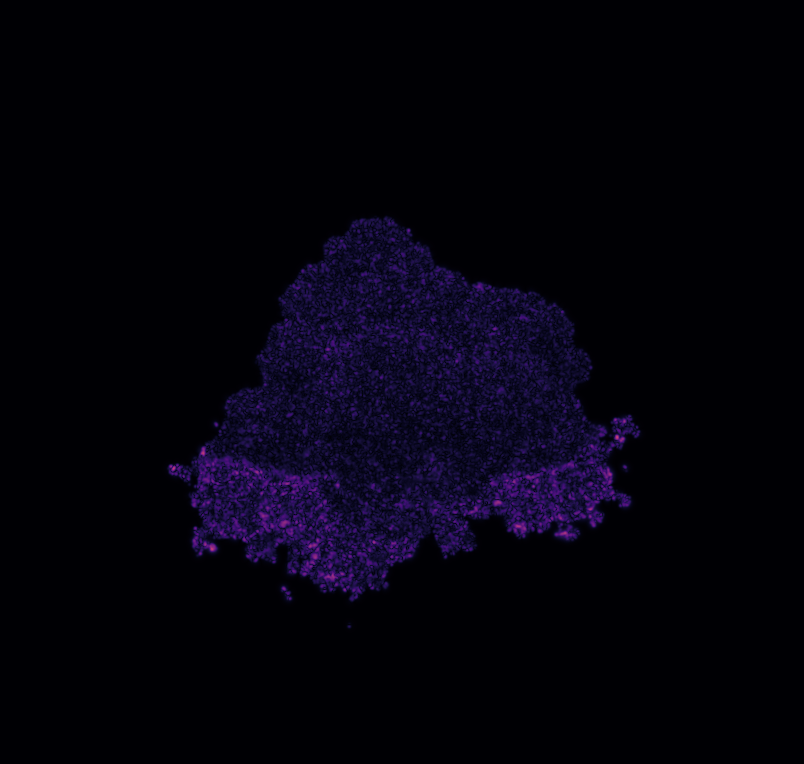}{\FLIP: 0.017}
  \teasertikz{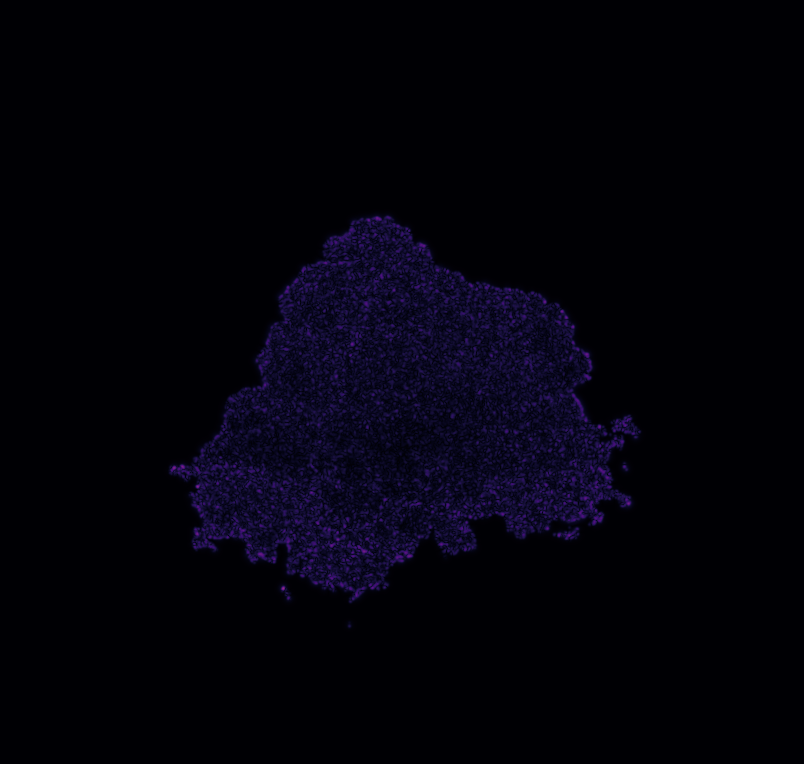}{\FLIP: 0.014}
  \teasertikz{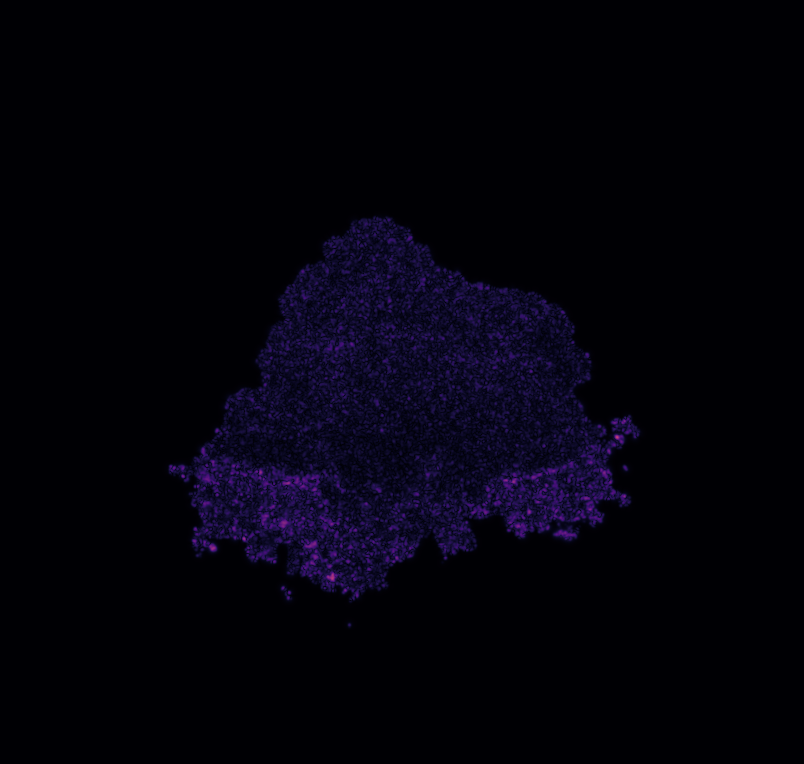}{\FLIP: 0.015}
\vspace{-2em}
  \caption{\label{fig:zfp-comparison-iso}
  Direct image comparison between our compression and ZFP \rev{(we use
  \emph{dense} textures to render the latter)}. The ZFP
  representation is obtained by compressing and directly decompressing the WDAS
  cloud, and rendering the resulting structured-regular volume with an implicit
  ISO surface ray marcher.  The color mapping shows reconstructed surface
  normals. We compare the outcome using \FLIP.}
\vspace{-2em}
\end{figure*}
We also compare \rev{the quality of our compression with that of} ZFP. We do not implement a full
ZFP renderer due to the technical debt involved. Instead, we compress and then
decompress the volume data sets from \cref{tab:datasets} with ZFP using
different compression rates. The decompressed volumes we compare to the
original, uncompressed data sets, allowing us to compute MSE and PSNR presented
in \cref{fig:zfp-comparison}.

We are also interested in a qualitative comparison of the two compression
algorithms given rendered images. For further analysis we implemented an
interval-based implicit ISO surface ray marcher to extract the $0.5$ ISO
surface of the WDAS cloud data set.  We show renderings with both our algorithm
as well as with a ZFP back-and-forth compressed and decompressed
structured-regular volume, in \cref{fig:zfp-comparison-iso}, as well as
difference images using \FLIP~\cite{flip}.

\subsection{VDB and Dense Texture Back-and-forth Conversion}
\begin{table}[th]
\begin{center}
  \setlength{\tabcolsep}{.6ex}
\begin{tabular}{r|c|c|c|c}
\toprule
Data Set                  & Original & Structured & Size MSE=0 & MSE (orig.\ size) \\
\midrule
WDAS $1 \times$           & 4.1~GB   & 26.3~GB    & 5.8~GB     & 0.001     \\
WDAS $\frac{1}{2}\times$  & 590~MB   & 3.3~GB     & 605~MB     & 0.010     \\
WDAS $\frac{1}{4}\times$  & 92.5~MB  & 413~MB     & 91.7~MB    & 0.000     \\
WDAS $\frac{1}{8}\times$  & 17~MB    & 52.2~MB    & 15.3~MB    & 0.000     \\
WDAS $\frac{1}{16}\times$ & 5.1~MB   & 6.7~MB     & 2.95~MB    & 0.000     \\

\bottomrule
\end{tabular}
\end{center}
\caption{\label{tab:backandforth}%
Back-and-forth conversion of the WDAS cloud's original VDB to dense texture and
back to our NanoVDB. We report the size of the original NanoVDB, the size of
the dense grid, the size of ours when targeting zero error (MSE=0), and MSE
when compressing the structured grid so the size is exactly that of the
original.
}
\vspace{-2em}
\end{table}
Disney's cloud data set~\cite{wdas-cloud} was originally published as VDB. For
our purposes, we converted it to a structured-regular format matching the exact
resolution of the original VDB. This allows us to compare the compressed size
as well as the quality achieved of our compression algorithm to that of the
original VDB. Conveniently, Disney's data repository contains the data set in
different sizes, each downsampling the original data set by a factor ranging
from $\frac{1}{2}$ to $\frac{1}{16}$ along each axis. This allows us to run
the experiment using different resolutions of the data set. As the original
format is (Open-)VDB, we convert the five data sets to NanoVDB first. The
NanoVDB file format allows the data to be compressed internally, either using
ZIP or BLOSC---non of which is useful to us as these formats do not support
random access. We are careful to deactivate internal compression during
conversion.

In \cref{tab:backandforth} we present the results of our experiments. We report
the size of the original data converted to NanoVDB and the size of
the corresponding structured-regular grid. We further report the smallest size
in bytes our algorithm achieves while compressing the data with zero error
(MSE=0). For the original resolution, e.g., we observe that compression without
error gives a size of 5.8~GB while the original data was 4.1~GB in size. We
finally also report the error we realize when compressing the data to the exact
size of the original; when compressing the high-resolution data set to match
the original size of 4.1~GB, e.g., the MSE we realize is $0.001$.

\section{Discussion}
We presented a fixed-rate compression algorithm to encode structured-regular
volumes with OpenVDB. Our main goal was to evaluate the fitness of VDB for
sci-vis data when used as a drop-in replacements for dense 3D textures in a
GPU-based volume path tracer. From an engineering perspective this is appealing
in many ways, as NanoVDB provides an interface that is very similar to that of
dense 3D textures, and is equally easy to use.

It was not the main goal of this work to devise a most efficient
hierarchical compression algorithm, in a sense that it is generally known that
hierarchical encoding has beneficial compression properties when applied to
volumes~\cite{star-compressed-vr}. Much more than that, properties of VDB such
as the worst case compression rate when applied to data that is not sparse at
all, or their sampling performance when compared to dense textures are much
more important to us; and we can conclude that even in adversarial cases, both
volume size as well as rendering performance are well within range of the
performance of just using dense textures.

With our VDB compression we note that the data should be inherently sparse. But
even if it is not, from an engineering perspective, using VDB still has
advantages as the size and performance overhead is so small. We however note
that sparseness should come from the volume itself and not (only)
from the transfer function. Compression is performed in a pre-process and the
VDB structure does not automatically adapt to topological changes only induced
by the transfer function. The volume must contain a value clearly
distinguishable as background, and with high compression rates, adversarial
transfer functions can still reveal block artifacts.

It is not our goal in this paper to reduce those artifacts, or to propose the
best possible compression algorithm. Optimizations like that are future work
and orthogonal to what we propose. Another interesting future work is
converting other data structures such as AMR or finite element meshes to VDB
using voxelization, to eventually replace all of them with VDB. More research
is necessary to determine if this is viable, \rev{since especially in the case
of AMR, level differences in the hierarchy often span multiple orders of
magnitude of space, whereas the standard VDB layouts support a fixed number of
consecutive hierarchy levels only.}

\section{Conclusion}
We proposed to use hierarchical encoding with OpenVDB and NanoVDB to compress
structured-regular 3D grids for volume rendering with Monte Carlo path tracing.
Our main goal was to evaluate the fitness of VDBs for sci-vis volume
rendering. We conclude that VDBs are viable alternatives to dense GPU textures
that incur only little memory and performance overhead even in adversarial
cases. For sparse data as used in our evaluation we can achieve high
compression rates at comparably high quality. \rev{Qualitatively,} our
fixed-rate compression algorithm, although much simpler, compares favorably to
compression with ZFP if the data is sparse. In the future we want to evaluate
if VDBs can also be used to encode other data types typically found in sci-vis,
such as finite elements or AMR grids.

\vspace{-0.5em}
\section*{Acknowledgments}
\ifsubmission
Intentionally omitted for review.
\else
This work was supported by the Deutsche Forschungsgemeinschaft (DFG, German
Research Foundation) under grant no.~456842964. This work was also supported by
the SPACE project under grant agreement No 101093441. The project is supported
by the European High-Performance Computing Joint Undertaking and its members
(including top-up funding by the Ministry of Education, Youth and Sports of the
Czech Republic ID: MC2304). This work was also supported by the Ministry of
Education, Youth and Sports of the Czech Republic through the e-INFRA CZ
(ID:90254).
\fi
\vspace{-0.5em}

\bibliographystyle{eg-alpha-doi}  
\bibliography{main}

\end{document}